\documentclass{jfm}

\usepackage{hyperref}
\hypersetup{
	colorlinks = true,
	allcolors = {blue}
}
\usepackage[justification=justified]{caption}
\usepackage{natbib}
\usepackage{amssymb}
\usepackage{amsmath}
\usepackage{array}
\usepackage[utf8]{inputenc}
\usepackage[dvipsnames,svgnames,table]{xcolor}
\usepackage{verbatim}
\usepackage{graphicx}
\usepackage{dcolumn}
\usepackage{bm}
\usepackage{booktabs}

\usepackage{siunitx}
\usepackage{tikz}
\usepackage{subfig}
\usepackage{makecell}
\usepackage{nicefrac}
\usepackage{nth}

\usepackage{makecell}

\usepackage{cleveref}
\crefname{figure}{figure}{figures} 
\Crefname{figure}{Figure}{Figures} 

\setcellgapes{3pt}

\newcommand{\vect}[1]{\ensuremath{{\bm{#1}}}}
\newcommand{\mat}[1]{\mathbf{#1}}
\newcommand{\grad}{\vect{\nabla}}
\newcommand{\gradperp}{\vect{\nabla}_\perp}
\newcommand{\xperp}{\vect{x}_\perp}

\newcommand{\deriv}[2]{\ensuremath{\frac{{\rm d}#1}{{\rm d}#2}}}
\newcommand{\ddt}[1]{\deriv{#1}{t}}

\newcommand{\partiald}[2]{\ensuremath{\frac{\partial#1}{\partial#2}}}
\newcommand{\partialdd}[2]{\ensuremath{\frac{\partial^2#1}{\partial#2^2}}}

\newcommand{\vperp}{\vect{v}_\perp}

\newcommand{\adim}[1]{{#1}}
\newcommand{\ddim}[1]{\tilde{#1}}

\newcommand{\heq}{{h_{\mathrm{eq}}}}

\newcommand{\Reyn}{{\mathrm R_{\mathrm e}}}
\newcommand{\Reynbv}{\mathrm{Re}_\mathrm{bv}}
\newcommand{\Reyneq}{\mathrm{Re}_\mathrm{eq}}

\newcommand{\Sq}{\mathrm S_{\mathrm q}}
\newcommand{\Sqbv}{\mathrm{Sq}_\mathrm{bv}}

\newcommand{\hbv}{\tilde H_{\rm bv}}

\newcommand{\OO}[1]{\mathcal{O}({#1})}
\newcommand{\heaviside}[1]{\mathbb H\left({#1}\right)}

\title{Viscous adhesion in vibrated sheets: elastohydrodynamics with inertia and compressibility effects}
\shorttitle{Viscous adhesion in vibrated sheets}

\author{
	Stephane Poulain\aff{1} \corresp{\email{stephapo@uio.no}},
	Timo Koch\aff{1,2},
	L. Mahadevan\aff{3,4}
	\and Andreas Carlson\aff{1,5}
	\corresp{\email{acarlson@uio.no}}
}
\shortauthor{S. Poulain, T. Koch, L. Mahadevan and A. Carlson} 
\affiliation
{
\aff{1}
Mechanics Division, Department of Mathematics, University of Oslo, 0316 Oslo, Norway
\aff{2}
Department of Scientific Computing and Numerical Analysis, Simula Research Laboratory, 0164 Oslo, Norway
\aff{3}
Paulson School of Engineering and Applied Sciences, Harvard University, Cambridge, MA 02138, USA
\aff{4}{
Department of Physics and Department of Organismic and Evolutionary Biology, Harvard University, Cambridge, MA 02138, USA
}
\aff{5}{
Department of Medical Biochemistry and Biophysics, Ume{\aa} University, 901 87, Ume{\aa}}, Sweden
}

\begin{document}

\maketitle
\begin{abstract}
Inspired by recent experiments demonstrating that vibrating elastic sheets can function as seemingly contactless suction cups, we investigate the elastohydrodynamic hovering of a thin elastic sheet vibrating near a rigid substrate.
Previous theoretical work suggests that the hovering height results from a balance between the active forcing that triggers the vibrations, the bending forces associated with the sheet's deformation, the viscous lubrication flow between the sheet and the substrate, and the sheet's weight.
Here, we extend this analysis beyond the asymptotic regime of weak forcing and explore the regime of strong forcing through numerical simulations.
We further quantify the influence of fluid inertia and compressibility on the equilibrium hovering height and the maximum load that can be supported. Both effects are found to introduce repulsive contributions to the net force on the sheet, which can significantly reduce its adhesive strength.
Beyond providing insights into soft contactless grippers and swimming near surfaces, our analysis is relevant to the elastohydrodynamics of squeeze films and near-field acoustic levitation.
\end{abstract}


\section{Introduction}
Contactless gripping or hovering near surfaces is desirable in many applications \citep{Vandaele2005}.
Fluid-mediated strategies include acoustic levitation to manipulate small objects \citep{Andrade2018}, Bernoulli grippers to grab delicate items \citep{Waltham2003,Li2015}, hovercrafts traveling on air cushions, and ground-effect flight used by both birds and vehicles
\citep{Rayner1991,Ollila1998}.
In parallel, soft robotics \citep{Whitesides2019} has emerged as an alternative to traditional mechanical systems with applications in handling fragile objects \citep{Shintake2018} and bio-inspired locomotion \citep{Calisti2017}.
Combining contactless dynamics with soft designs, \cite{Argentina2007} theoretically proposed that an elastic sheet sustaining traveling waves and placed next to a surface could levitate and translate owing to elastohydrodynamic interactions.
More recently, \citet{Colasante2015} and \cite{Weston2021} devised a novel strategy: they showed experimentally that attaching a vibration motor to an elastic sheet creates a seemingly contactless suction cup that is able to adhere to surfaces or pick up objects weighing up to several kilograms (see \cite{ColasanteYT} and references in \cite{Ramanarayanan2024}), offering a possible alternative to state-of-the-art contactless grippers.
In our earlier publication \citep{Poulain2024}, we modeled the viscous flow in the thin gap between a wall and an actuated elastic sheet.
We demonstrated how a time-reversible forcing of the soft sheet triggers a non-reversible response, an effect previously examined in the context of microorganism swimming with flagella \citep{Wiggins1998b,Wiggins1998a,Yu2006,Lauga2007}.
Indeed, elastohydrodynamic interactions enable breaking the time-reversal symmetry of viscous flows and circumventing the scallop theorem \citep{TaylorMovie,Purcell1977,Bureau2023,Rallabandi2024}, generating a net effect that attracts or repels the sheet depending on the forcing's spatial profile.
For a localized central forcing, we showed that the elastohydrodynamic effects attract the sheet towards the surface against gravity, enabling adhesion and hovering.
This viscous elastohydrodynamic mechanism rationalizes qualitatively the experiments of \cite{ColasanteYT} and \cite{Weston2021}.
While our analysis isolated the essential physics, it relied on asymptotic calculations assuming a weak forcing and on the assumptions of incompressible, inertialess flow, which builds on lubrication theory. A more comprehensive understanding, however, may require considering the inertia and compressibility of the fluid, as highlighted by \citet{Ramanarayanan2022,Ramanarayanan2022b,Ramanarayanan2024}.

Corrections to lubrication theory have long been studied, particularly in squeeze-film settings relevant to bearings \citep{Moore1965} and resonant micro-electro-mechanical systems (MEMS) \citep{Bao2007,Pratap2014,Fedder2015}.
In standard squeeze films, two surfaces of length $\tilde R$ immersed in a fluid with ambient pressure $\tilde p_a$, ambient density $\tilde \rho_a$ and dynamic viscosity $\tilde \mu$  oscillate with the distance separating them evolving harmonically as $\tilde h(\tilde t)=\tilde h_0\left(1+a\sin(\tilde \omega \tilde t)\right)$, where $0<a<1$.
Experimental observations reveal that the flow in the gap can generate net normal forces or damping effects that are not predicted by classical lubrication theory, which only considers viscous effects.

Considering the viscous and compressible flow of an ideal gas, \cite{Taylor1957} and \citet{Langlois1962} showed that a repulsive normal force between the two surfaces arises for finite Squeeze number ${\rm Sq}=\tilde \mu \tilde \omega \tilde R^2/\tilde h_0^2 \tilde p_a>0$, i.e., when the magnitude of viscous stresses is comparable to the ambient pressure.
These effects, studied in more detail since \citep{Bao2007,Melikhov2016,Ramanarayanan2022}, allow for the near-field acoustic levitation of small objects \citep{Shi2019}, are relevant to MEMS sensors such as atomic force microscopes employing vibrating micro-cantilever plates or beams \citep{Bao2007,Wei2021}, and may have applications in the design of haptic surfaces \citep{Wiertlewski2016}.
More generally, squeeze films belong to an interesting class of elastohydrodynamic problems where compressible effects are significant even at low Mach numbers (e.g., \citet{Mandre2009,Peng2023}).
Compressible squeeze flows can also be coupled with elastic deformations.
Instead of using rigid-body vibrations to induce near-field acoustic levitation, leveraging flexural vibrations has been proposed to potentially enhance levitation efficiency \citep{Hashimoto1996,Minikes2003} and allow lateral translation of the levitated object \citep{Ueha2000,Andrade2018}. 
The coupling between the elastic deformations of thin cantilever plates with squeeze flows is also key for MEMS \citep{Lee2009,Pandey2007}.

Fluid inertia can also contribute to squeeze film dynamics, and inertial corrections to lubrication theory have been derived for finite  Reynolds number ${\rm Re}=\tilde \rho_a \tilde \omega \tilde h_0^2/\tilde \mu>0$ in the context of bearings \citep{Ishizawa1966,Kuzma1968,Tichy1970,Jones1975} and later applied to acoustic levitation \citep{Atalla2023, Liu2023} and MEMS \citep{Veijola2004}.  This corresponds to the regime where the inertial and diffusive timescales are comparable.
For incompressible flows, inertia leads to a repulsive normal force between the surfaces.
Beyond applications to bearings, inertial effects are also important in near-field acoustic levitation 
More broadly, flows at intermediate Reynolds numbers, where both viscous and inertial effects are significant, arise in a variety of physical systems.
Notable examples include inertial corrections to pulsatile flows in tubes relevant to physiological systems \citep{Womersley1955}, swimming of small organisms \citep{Gazzola2014,Derr2022}, and oscillatory flows around spheres motivated by Atomic Force Microscopes \citep{Fouxon2018,Fouxon2020,Zhang2023} and Surface Force Apparatus \citep{Bigan2024}.
In confined environments, the aforementioned analytical corrections to Reynolds' lubrication theory are limited to rigid geometries or specific deformations. 
\citet{Rojas2010} derived lubrication equations with inertial corrections which naturally accommodate arbitrary deformable geometries. They have successfully applied this framework to free-surface phenomena; however, to our knowledge, it has not yet been extended to elastohydrodynamics and soft lubrication \citep{Skotheim2005}.

Interestingly, an object may experience a net attractive force towards a vibrating surface rather than a repulsive force, an effect leading to inverted near-field acoustic levitation \citep{Takasaki2010,Andrade2020}.
This observation has recently been theoretically rationalized by \citet{Ramanarayanan2022}, who found that, surprisingly, incorporating both inertial and compressible effects in the lubrication dynamics reveals the possibility of an attractive force when including second-order inertial effects.
\cite{Ramanarayanan2022b,Ramanarayanan2024} later extended their analysis to deformable geometries, showing an enhancement of the attractive effect.
In contrast, our previous work \citep{Poulain2024} demonstrated that viscous, inertialess, and incompressible fluid-structure interactions alone can produce an adhesive effect when an elastic sheet is driven near a surface. In this regime, which we believe to be relevant to the experiments of \citet{Colasante2015} and \citet{Weston2021}, we anticipate that inertial and compressible effects play a secondary role, entering as corrections rather than dictating the primary viscous adhesion mechanism.
Indeed, near-field acoustic levitation typically supports only objects weighing a few milligrams \citep{Andrade2020}, whereas the viscous mechanism we uncovered predicts a lift capacity on the order of kilograms under typical experimental conditions, in agreement with observations.

In this article, we extend the results of \cite{Poulain2024} to study the adhesion of a forced elastic sheet to a substrate.
The system and governing equations are presented in \S \ref{sec:governing}.
We first revisit our previous results and study the system numerically beyond the asymptotic regime of weak forcing using classical viscous lubrication theory in \S \ref{sec:viscous}.
We then discuss independently the effects of fluid inertia in \S \ref{sec:inertia} and the effect of fluid compressibility in \S \ref{sec:compressible}.
We conclude with a discussion of the results in \S \ref{sec:conclusion}.

\section{Setup and governing equations}
\label{sec:governing}

\subsection{Setup}
\label{sec:governing_setup}

\begin{figure}
	\centering
\includegraphics[width=\linewidth]{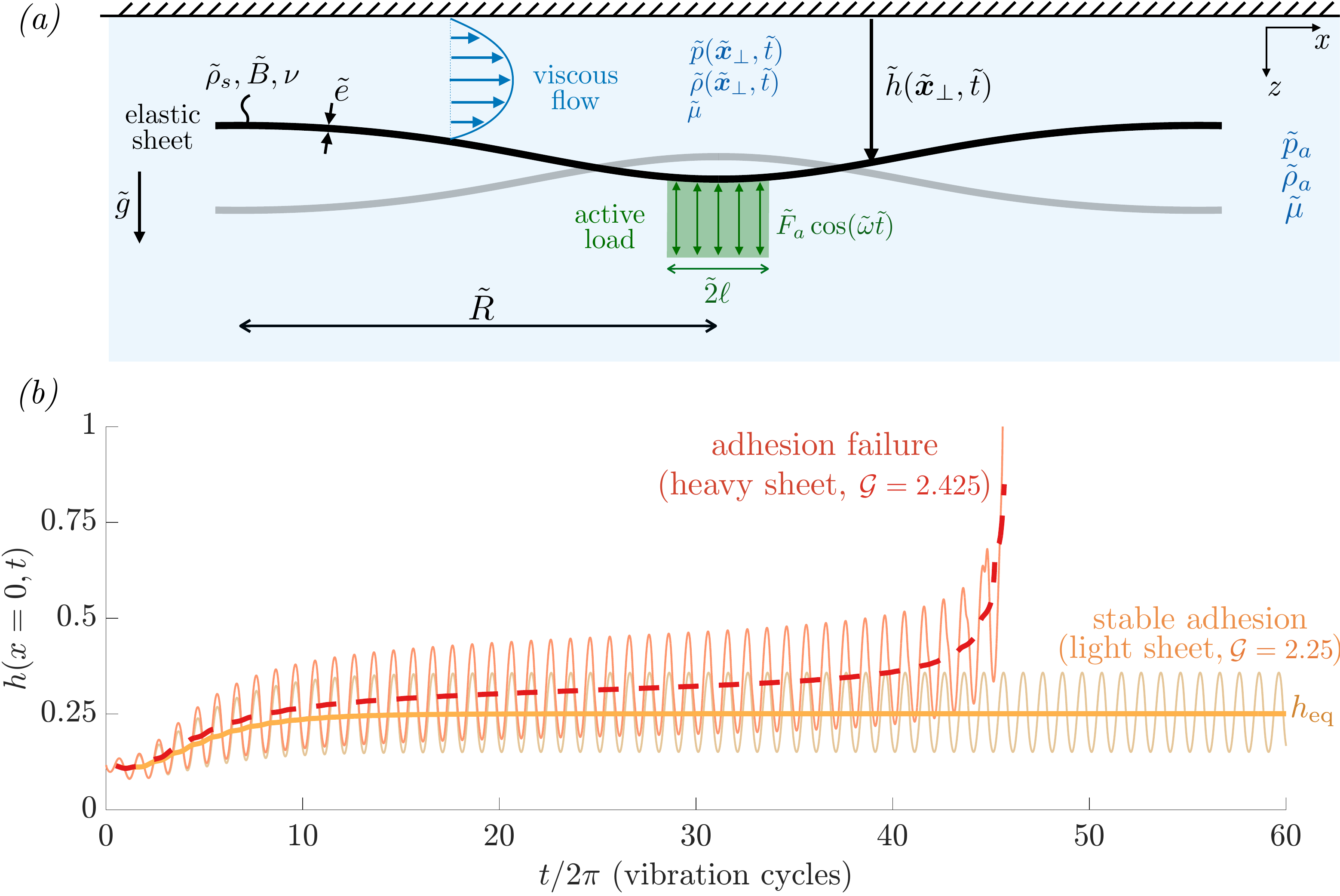}
	\caption{
    $(a)$
    An elastic sheet (radius $\tilde R$, density $\tilde \rho_s$, bending rigidity $\tilde B$, Poisson's ratio $\nu$, thickness $\tilde e$) immersed in a fluid (ambient density $\tilde \rho_a$, ambient pressure $\tilde p_a$, dynamic viscosity $\tilde \mu$) and forced periodically at its center (force $\tilde F_a$, angular frequency $\tilde \omega$, radius $\tilde \ell$) is placed below a solid substrate with gravity pointing downward. $\xperp=(x,y)$ represent the horizontal coordinates.
    $(b)$ When the dimensionless weight of the sheet $\mathcal G$ is small enough, the sheet hovers around an equilibrium position $h_{\rm eq}$ (illustrated here for $\alpha=5$, $\mathcal G=2.25$ and a purely viscous dynamics, $\mathcal I_{\rm bv}={\rm Re}_{\rm bv}={\rm Sq}_{\rm bv}=0$).
    Above a critical weight, the sheet cannot adhere to the substrate (shown for $\alpha=5$, $\mathcal G=2.425$). Thin lines represent the dimensionless gap thickness at the center of the sheet $h(x=0,t)$, while thick lines represent the time-averaged height $\langle h \rangle(x=0,t)$.
	\label{fig:schematic}}
\end{figure}

We consider the system shown in \cref{fig:schematic}: an elastic sheet of radius $\tilde R$, thickness $\tilde e$, density $\tilde \rho_s$, Poisson ratio $\nu$, Young's modulus $\tilde E$, bending modulus $\tilde B= \tilde E \tilde e^3/12(1-\nu^2)$, placed near a solid substrate.
The surrounding fluid is Newtonian with ambient density $\tilde \rho_a$, ambient pressure $\tilde p_a$, and a constant viscosity $\ddim \mu$.
The sheet is forced at its center by a harmonic active load with angular frequency $\tilde \omega$, radius $\tilde \ell$, and force magnitude $\tilde F_a$.
In this setting, we aim to establish equilibrium conditions under which the elastic sheet hovers at a finite, stable time-averaged distance from the wall despite the gravitational pull.

An important characteristic scale of the system is the elastohydrodynamic height scale  ~\citep{Poulain2024}
\begin{subequations}
\begin{align}
\hbv={\tilde R^2}{\left(\frac{\tilde \mu \tilde \omega}{\tilde B}\right)^{1/3}}, \label{eq:Hbv}
\end{align}
for which viscous stresses scaling as $\tilde \mu \tilde \omega \tilde R^2 / \hbv^2$ and bending stresses scaling as $\tilde B \hbv / \tilde R^4$ balance. Associated with $\hbv$ is the aspect ratio $\varepsilon_{\rm bv}=\hbv/\tilde R$, and a elastohydrodynamic force scale 
\begin{align}
    \tilde F_{\rm bv} = (\tilde \mu \tilde \omega \tilde B^2 )^{1/3},
\end{align}
\end{subequations}
with $\tilde F_{\rm bv}/\tilde R^2$ the scale for elastohydrodynamic stresses.
In the remainder of the article, we work with dimensionless quantities (written throughout without a tilde, in contrast to dimensional quantities):
\begin{align}
\begin{split}
    \adim t = \ddim t\ddim\omega, \quad
    {\vect x}_\perp=\frac{\tilde{\vect x}_\perp}{\ddim R}, \quad  
    h(\xperp,t)=\frac{\tilde h(\tilde{\vect x}_\perp,\tilde t)}{\hbv}, \\
\adim p(\xperp,t) = \frac{\ddim p(\tilde{\vect x}_\perp,\tilde t)}{\tilde F_{\rm bv}/\tilde R^2},\quad
\rho(\xperp,t)=\frac{\ddim \rho(\tilde{\vect x}_\perp,\tilde t)}{\ddim \rho_a},
\end{split}
\end{align}
with $\xperp=(x,y)$ the horizontal coordinates, $p(\xperp,t)$ the fluid pressure relative to the ambient pressure, and $\rho(\xperp,t)$ the fluid density. 

We consider small aspect ratios, $\varepsilon_{\rm bv} \ll 1$, for which the viscous flow in the thin layer separating the sheet and the wall dominates the system dynamics.
Two other dimensionless numbers characterize the flow.
Inertial effects are characterized by the film Reynolds number $\Reynbv=\tilde \rho \tilde \omega \hbv^2/\tilde \mu$ that compares the inertial pressure $\tilde \rho \tilde R^2 \tilde \omega^2$, with $\tilde \omega \tilde R$ the scale for the horizontal velocity, to the elastohydrodynamic viscous stress $\tilde F_{\rm bv}/\tilde R^2$ \citep{Batchelor}.
The Reynolds number can also be written as ${\rm Re}_{\rm bv}=(\tilde H_{\rm bv}/\tilde \delta)^2$, with $\tilde \delta=(\tilde \mu/\tilde \rho \tilde \omega)^{1/2}$ the viscous penetration length, the lengthscale for diffusion of vorticity: in other contexts, this Reynolds number may also be referred to as a Womersley number.
Compressible effects are characterized by the Squeeze number $\Sqbv = \tilde F_{\rm bv}/\tilde p_a \tilde R^2$ that compares the viscous stress to the ambient pressure $\tilde p_a$ \citep{Taylor1957}.
With \eqref{eq:Hbv}, the inertial and compressible effects in the fluid are  characterized respectively by
\begin{align}
    \Reynbv=\frac{\ddim \rho_a \ddim \omega^2 \ddim R^4 }{\left(\ddim \mu \ddim \omega \ddim B^2\right)^{1/3}}, \qquad
    \Sqbv=\frac{\left(\tilde \mu \tilde \omega \tilde B^2\right)^{1/3}}{\tilde p_a \tilde R^2}.
    \label{eq:dim1}
\end{align}

\subsection{Inertial lubrication}
Let us find a depth-integrated description of the flow in the thin gap between the sheet and the wall ($\varepsilon_{\rm bv}\ll 1$) which includes the first-order effects of inertia at $\OO{\Reynbv}$ and compressibility at $\OO{\Sqbv}$.
Mass conservation yields
\begin{subequations}
\begin{align}
\partiald{\left(\rho h\right)}{t} +  \grad_\perp \cdot\left(\rho \vect q\right) &= 0, 
\label{eq:massbalance}
\\
\rho &= 1 + \Sqbv p,\label{eq:perfectgas}%
\end{align}\label{eq:massbalance2}%
\end{subequations}%
with $\vect q=\int_0^h \vperp {\rm d}z=\tilde{\vect q} / \tilde \omega \tilde L $ the horizontal volumetric fluid flux, $\vperp$ the horizontal fluid velocity profile, and $\gradperp=(\partial/\partial x, \partial/\partial y)$ the horizontal gradient operator.
\Cref{eq:perfectgas} is the dimensionless ideal gas law assuming isothermal conditions, $\tilde \rho/\tilde \rho_a=1+\tilde p/\tilde p_a$, an assumption we discuss later in \S \ref{sec:compressible}.
The volumetric flux $\vect q$ appearing in \eqref{eq:massbalance2} is found from the Navier-Stokes equations. \Citet{Rojas2010} describe a procedure to consider the first-order inertial corrections to lubrication theory for free surface flows with $\varepsilon_{\rm bv}\ll 1$, $\Reynbv \ll 1$.
We adapt their derivation to a fluid layer bounded by two solid walls (\cref{sec:derivationNS}), which yields the depth-integrated horizontal momentum balance, to $\mathcal O(\varepsilon_{\rm bv}^2,\Reynbv,\varepsilon_{\rm bv}^2\Reynbv,\Reynbv\Sqbv,\Sqbv)$:
\begin{align}
    12   {\vect q}+ { h^3}{\grad}_\perp  p + \Reynbv h^3 \left[
    \frac65 \partiald{}{ t}\left(\frac{{\vect q}}{ h}\right)
     +\frac{54}{35} \frac{{\vect q} }{ h} \cdot \grad_\perp\left(\frac{{\vect q}}{ h}\right)
     - \frac6{35} \frac{{\vect q}}{ h^2}\partiald{ h}{ t}
    \right]&=0.
    \label{eq:fluxRe}
\end{align}
The first two terms in  \eqref{eq:fluxRe} recover the Reynolds equation from classical lubrication theory, valid for vanishing Reynolds number \citep{Batchelor} and linking volumetric flux to pressure gradients.
The term proportional to $\Reynbv$ corresponds to the first-order correction due to fluid inertia. It corrects the parabolic Poiseuille velocity profile of classical lubrication theory and predict a sextic velocity profile, discussed in \S\ref{sec:inertia}.
\cite{Rojas2010} found good agreement between experiments and this extended lubrication theory for Reynolds numbers of order one.
Previous theoretical studies from  \citet{Ishizawa1966} and \citet{Jones1975}, albeit limited to non-deformable boundaries ($h$ independent of $\xperp$), suggest that this first-order correction may even be valid as long as the Reynolds number is less than $100$. \Cref{eq:fluxRe} recovers these studies and generalizes them to an arbitrary height $h(\xperp, t)$, suggesting that \eqref{eq:fluxRe} may be valid for finite, relatively large, values of $\Reynbv$, say $\OO{10}$.

We emphasize that both the unsteady and nonlinear advective terms scale with $\Reynbv$ and note that $\vect q/h$ is the averaged horizontal velocity in the gap and $(1/h) \partial h/\partial t$ the average vertical gradient of vertical velocity, showing the connection between the inertial corrections in \eqref{eq:fluxRe} and the acceleration terms in the Navier-Stokes equations. 
This contrasts with systems in which objects oscillate rapidly in a fluid but with small amplitude.
In such cases, inertial effects are dominated by the unsteady acceleration, which scales as the oscillation amplitude, while convective terms remain negligible because they scale as its square.
As a result, the unsteady Stokes equation suffices to capture the flow and resulting forces  \citep{Fouxon2018,Fouxon2020,Zhang2023,Bigan2024}.
In our case, however, the amplitude of the sheet's oscillation can become comparable to the gap thickness, justifying the inclusion of all inertial terms in the analysis.

\subsection{Elastic deformations}
To close the system formed by \eqref{eq:massbalance2} and \eqref{eq:fluxRe}, the fluid pressure $p$ must be linked to elastic stresses by considering a vertical momentum balance of the elastic sheet. 
We here adopt the Kirchhoff-Love model \citep{Timoshenko1959,Landau1986} to describe the sheet's deformation.
This model neglects any stretching of the sheet, which is justified for a one-dimensional sheet that deforms cylindrically (in which case the model reduces to the Euler-Bernoulli beam model) and for two-dimensional sheets with deformations that remain small compared to the sheet's thickness.
The normal force per unit area from the aerodynamic interactions in the thin gap is  $p + \OO{\varepsilon_{\rm bv}^2}$, so that the normal force balance in the thin-film limit reads:
\begin{equation}
\begin{split}
	 \mathcal I_{\rm bv} \partialdd{h}{t} &=  p + \grad_\perp \cdot \left( \grad_\perp \cdot \mat M\right) +  f_a(\xperp,t)+\mathcal G + f_{w}(h), \\
	 \mat M &=-\left[\left(1-\nu\right) \boldsymbol \kappa + \nu \operatorname{tr}(\boldsymbol \kappa) \right].
\end{split}
\label{eq:normalForceBalance_dimensionless}
\end{equation}
The dimensionless number $\mathcal I_{\rm bv}=\tilde \rho_s \tilde e {\tilde H}_{\rm bv} \tilde \omega^2 \tilde R^2/ \tilde F_{\rm bv}$ compares solid inertia to the elastohydrodynamic stress scale. The right-hand-side of \eqref{eq:normalForceBalance_dimensionless} corresponds respectively to the stress from the fluid, the bending stress (with $\boldsymbol \kappa$ the Hessian of $h$---the curvatures of the sheet), the periodic active stress $f_a$, the sheet's areal weight with $\mathcal G=\tilde \rho_s \tilde e \tilde R^2 \tilde g/\tilde F_{\rm bv}>0$, and a term preventing collision with the wall. We also note that in the case of a constant bending rigidity considered here, the bending stresses simplify to $\grad_\perp \cdot \left( \grad_\perp \cdot \mat M\right)=-\nabla^4_\perp h$.

The active forcing behind $f_a$ is harmonic in time and is distributed at the center of the sheet with a dimensionless radius $\ell=\ddim l/\ddim R$ (\cref{fig:schematic}):
 \begin{align}
    f_a(\xperp,t)=\alpha\cos(t) \frac{1-\mathbb H(\lvert \xperp\rvert-\ell)}{\ell},
    \label{eq:activeforcing}
\end{align}
with $\mathbb H$ the Heaviside function and $\alpha=\tilde F_a/\tilde F_{\rm bv}$ the relative strength of the forcing compared to the elastohydrodynamic force scale.
With \eqref{eq:Hbv}, the dimensionless numbers in \eqref{eq:normalForceBalance_dimensionless} read
\begin{align}
\begin{split}
	\mathcal I_{\rm bv}&=\frac{\tilde \rho_s \tilde e \tilde \omega^2 \tilde R^4}{\tilde B} , \quad
    \mathcal G = \frac{\tilde F_g}{\tilde F_{\rm bv}} = \frac{\tilde \rho_s \tilde e \tilde g \tilde R^2}{\left(\ddim \mu \ddim \omega \ddim B^{2}\right)^{1/3}}, \quad
    \alpha=\frac{\tilde F_a}{\tilde F_{\rm bv}}=\frac{\ddim F_a}{\left(\ddim \mu \ddim \omega \ddim B^{2}\right)^{1/3}}.
\end{split}
\label{eq:dim2}
\end{align}

Finally, $f_w(h)$ models elastic collisions between the sheet and the wall.
As discussed later in \S \ref{sec:contactless}, we observe numerically that when forcing and weight both exceed a critical value, the edges of the sheet may contact the wall. To handle these collisions numerically, we add a local repulsive force $f_w(h)=\left(A/h\right)^n$, with $n \geq 1$, modeling an elastic collision with a contact that effectively occurs at the dimensionless height $A\ll 1$.

 \subsection{Boundary conditions}

\label{sec:boundaryConditions}
The edges of the elastic sheet are free of bending moment, twisting moment, and shear force. The appropriate boundary conditions are~\citep[p.~586]{Naghdi1973}:
\begin{align}
    \mat M \vect{e_r} \cdot \vect{e_r}= 0, \quad
    \left(\gradperp \cdot \mat M \right)\cdot \vect{e_r} + \gradperp\left( \mat M \vect{e_r} \cdot \vect{e_\theta}\right)\cdot \vect{e_\theta}=0,
    \label{eq:bc_elastic}
\end{align}
where $\vect{e_r}$ is the outward unit normal at the edge of the sheet and $\vect{e_\theta}$ is the tangent.
As the third and final boundary condition, we set the pressure at the edges of the sheet as 
\begin{align}
    p=
    \begin{cases}
     0~&\text{if}~\vect q \cdot \vect e_r > 0~~\text{(outflow)} \\
     - \dfrac{k}{2} \Reynbv\left(\dfrac{\vect q \cdot \vect e_r }{h}\right)^2~&\text{if}~\vect q \cdot \vect e_r < 0 ~~\text{(inflow)}
    \end{cases},
    \label{eq:bcPressureMain}
\end{align}
with $k=1/2$, a loss coefficient.
This boundary condition models the pressure loss when fluid enters the thin gap; it is justified and discussed in more detail in \Cref{sec:Pressurebc}.

\subsection{Numerical model}
We have formulated the governing equations \eqref{eq:massbalance}, \eqref{eq:fluxRe}, \eqref{eq:normalForceBalance_dimensionless} and boundary conditions \eqref{eq:bc_elastic} and \eqref{eq:bcPressureMain} for a 2-dimensional (2D) sheet for completeness. 
For purely viscous flows, we have shown \citep{Poulain2024} that there are no qualitative differences between 1D and 2D flows. Accordingly, we restrict the present study to 1D for simplicity.
Hence, we consider $\xperp \rightarrow x$ and $\partial/\partial y=0$.
We solve the governing equations in conservative form
\begin{subequations}
\begin{align}
	\partiald{\boldsymbol{S}}{t} +
	 \partiald{\boldsymbol{F}}{x} = \boldsymbol{Q},
\end{align}
with
\begin{equation*}
\begin{array}{@{} >{{}}r<{{}} @{}r@{\,} *{5}{c} @{\,}l@{}}
\boldsymbol{S}(\boldsymbol{U}) = & [ & \rho h,       & \mathcal I_{\rm bv} v,       & 0,          & \dfrac{6\Reynbv}{5}u,      & h     & ], \\[2.5ex]
 \boldsymbol{F}(\boldsymbol{U})  = & [ & 0, & -\dfrac{\partial m}{\partial x}, & -\dfrac{\partial h}{\partial x}, & \dfrac{27}{35}\Reynbv u^2 + p, & hu & ], \\[2.5ex]
 \boldsymbol{Q}(\boldsymbol{U}) = & [ &\rho v,       & p + f_a(x,t)+\mathcal G + f_w(h),       & m,          & -\dfrac{12u}{h^2} + \dfrac{6}{35}\Reynbv\dfrac{uv}{h},      & 0     & ], \\
\end{array}
\end{equation*}
and the five primary unknowns $\boldsymbol{U}= [v, m, h, u, p]$,
representing the vertical sheet velocity $v={\partial h}/{\partial t}$, bending moment $m=- \partial^2 h/\partial x^2$, height $h$, average horizontal fluid velocity $u=q/h$, and fluid pressure $p$, respectively. We note that $\rho=1+\Sqbv p$.

We consider a domain $x \in [0,1]$ with symmetric boundary conditions at $x=0$ and the appropriate boundary conditions at $x=1$:
\begin{align}
m = 0, \quad \partiald mx = 0, \quad p&=
\begin{cases}
 0~&\text{if}~u \geq 0, \\
 - k\Reynbv u^2/2~&\text{if}~u  < 0.
\end{cases}
\end{align}
\label{eq:numericsEq}
\end{subequations}

We solve \eqref{eq:numericsEq} using the DuMu$^x$ library~\citep{Koch2021}.
We discretize space using a staggered finite volume scheme, where pressure unknowns are located at cell centers and other unknowns are located at vertices. This setup avoids checkerboard oscillations in the fluid pressure.
The flux term in $u^2$ is treated with a first-order upwind scheme.
The equations are advanced in time using a diagonally-implicit $\nth{3}$-order Runge-Kutta scheme~\cite[Thm.~5]{Alexander1977}, and the nonlinear system at each Runge-Kutta stage is solved with Newton's method.
Lower-order methods either yielded unsatisfactory accuracy or required excessively small time step sizes.
We used time step sizes $10^{-4} \leq \Delta t \leq 0.2$ and spatial step sizes $0.002\leq \Delta x \leq 0.02$.
The numerical simulations have been run to a time-averaged steady state---which could take from $t=\OO{10}$ up to $t=\OO{10^6}$ depending on the parameters---or until the height diverged.
We systematically ensured that any divergence of the numerical solution was independent of the numerical parameters and therefore corresponded to adhesion failure.
For initial conditions, we considered a flat sheet, $\boldsymbol{U}(x,t=0)=[0, 0, h(x,t=0), 0, 0]$, with $h(x,t=0)$ a constant.
For large values of $\alpha$ and $\mathcal G$, this initialization sometimes lead to divergence, even though a time-averaged steady state exists.
In these cases, we initialized the simulation using the steady-state solution from a run with the same $\alpha$ but smaller $\mathcal{G}$ (numerical continuation).
The repulsion potential was either turned off ($A=0$) or, when needed, chosen as $f_w(h)=\left(A/h\right)^n$ with $A=10^{-5}$ and $n=5$. We have verified that this choice does not significantly affect the results as long as $A$ is small.

\subsection{Choice of dimensionless parameters}
\label{sec:equili}
\Cref{eq:massbalance2,eq:fluxRe,eq:normalForceBalance_dimensionless} depend on five dimensionless numbers defined in \eqref{eq:Hbv} and \eqref{eq:dim2}, with $\varepsilon_{\rm bv}$ not appearing in the governing equations and $\ell$ defined in \eqref{eq:activeforcing}. They are summarized in \cref{tab:dimless}.
Since dimensional quantities such as the sheet's bending rigidity $\tilde B$ or the excitation frequency $\tilde \omega$ enter multiple dimensionless groups, independently varying them in experiments is not feasible.
Nevertheless, numerical simulations enable us to disentangle the respective roles of the dimensionless numbers in the dynamics and to clarify the underlying physical mechanisms they influence.
We consider a uniform sheet with a uniformly distributed weight: $\mathcal I_{\rm bv}$ and $\mathcal G$ are constants. We also consider the limit where the forcing localized at a single point, $\ell \rightarrow 0$; in practice, we set $\ell=0.05$ in our numerical simulations. 
The effect of a finite $\ell$ and a sheet locally rigid at its center has already been discussed in prior work \citep{Poulain2024}.

To guide our study, we consider the experiments of \cite{Weston2021}, who report a time-averaged equilibrium height (\cref{fig:schematic}$b$) $\tilde h_{\rm eq} \approx 600 \unit{\micro\meter}$  for a sheet of thickness $\tilde e \approx 300 \unit{\micro\meter}$, Young's modulus $\tilde E \approx 3 \unit{\giga\pascal}$, radius $\tilde R \simeq 10 \unit{\centi\meter}$, vibrating at frequency $\tilde \omega = 2\pi \times 200 \unit{\hertz}$  and supporting a weight $\tilde W \approx 5\unit{\newton}$.
The density ratio between the sheet and air is $\tilde \rho_s/\tilde \rho_a\approx 10^3$.
The vibrations are generated by an eccentric rotating mass motor, with an estimated mass $\tilde m \approx 0.6 \unit{\gram}$ and gyration radius $\tilde r \approx 1 \unit{\milli\meter}$, yielding a driving force $\tilde F_a=\tilde m \tilde r \tilde \omega^2$.
This gives a dimensionless forcing strength $\alpha=\tilde F_a/\tilde F_{\rm bv}\approx 90$.
While this is overestimated due to the rigid central support \citep{Poulain2024}, it nevertheless suggests that the regime $\alpha=\OO{10}$ is relevant experimentally.

Although we consider a dynamics dominated by viscous forces, we are interested in corrections due to additional physical effects. 
From the above parameters, the heightscale is $\tilde H_{\rm bv} \approx 14 \unit{\milli\meter}$, while the dimensionless equilibrium height is only a small fraction of this value: $h_{\rm eq}=\tilde h_{\rm eq}/\tilde H_{\rm bv} \approx 0.04$.
Our theoretical and numerical analysis will recover this observation.
However, this shows that the dimensionless numbers based on $\tilde H_{\rm bv}$, such as $\Reynbv$, $\Sqbv$, and $\mathcal I_{\rm bv}$, which allow for a compact theoretical description of the system dynamics, are not accurate indicators of the relative effect of fluid inertia, fluid compressibility, or solid inertia, respectively, as compared to viscous effects when the system is at equilibrium height.
Hence, we additionally define dimensionless numbers using $\tilde h_{\rm eq}$ as the vertical scale:
\begin{align}
    \Reyneq = \frac{\tilde \rho_a \tilde \omega  \tilde h_{\rm eq}^2}{\tilde \mu}=h_{\rm eq}^2 \Reynbv, \quad
    {\rm Sq}_{\rm eq}=\frac{\tilde \mu \tilde \omega \tilde R^2}{\tilde h_{\rm eq}^2 \tilde p_a} = h_{\rm eq}^{-2} \Sqbv, \quad 
    \mathcal I_{\rm eq}=\frac{\tilde \rho_s\tilde e \tilde h_{\rm eq}^3 \tilde \omega}{\tilde \mu \tilde R^2}=h_{\rm eq}^{3}\mathcal I_{\rm bv}.
\end{align}
Unlike the original dimensionless groups, these quantities cannot be computed a priori since $\tilde h_{\rm eq}$ is selected by the system and is initially unknown.
The experimental values yield $\alpha=\OO{10}$,
and we systematically vary $\alpha$ in \S \ref{sec:viscous}.
The Reynolds number is $\Reyneq \approx 30$, indicating that fluid inertia may play a significant role, studied in \S \ref{sec:inertia}.
The parameter controlling solid inertia, however, is smaller, $\mathcal I_{\rm eq} \approx 0.6$, suggesting a weaker yet potentially non-negligible effect of solid inertia.
In this study, we focus on the influence of fluid effects and leave a systematic investigation of solid inertia for future work. A related investigations into the coupling between solid inertia within elastohydrodynamic systems can be found in \citet{Ramanarayanan2024}. 
Finally, although ${\rm Sq}_{\rm eq}\approx 0.006$, we show in \S\ref{sec:compressible} that compressibility can still noticeably affect the dynamics even for small Squeeze numbers.

\begin{table}
\renewcommand{\arraystretch}{2.5}  
\centering
\caption{Characteristic scales and dimensionless parameters.}
\label{tab:dimless}
\begin{tabular}{l l l r}
Symbol & Definition & Physical meaning & Unit \\
\midrule
$\tilde H_{\rm bv}$ & \makecell{$\displaystyle \tilde R^2 \left(\frac{\tilde \mu \tilde \omega}{\tilde B}\right)^{1/3}$} & height balancing viscous and bending effects & \si{\m} \\
$\tilde F_{\rm bv}$ & \makecell{$\displaystyle \left(\tilde \mu \tilde \omega \tilde B^2\right)^{1/3}$} & force balancing viscous and bending effects & \si{\newton} \\
$\varepsilon_{\rm bv}$ & \makecell{$\displaystyle {\tilde H_{\rm bv}}/{\tilde R}$} & aspect ratio & - \\
$\alpha$ & \makecell{$\displaystyle {\tilde{F}_a}/{\tilde F_{\rm bv}}$} & forcing amplitude/elastohydrodynamic & - \\
$\mathcal G$ & \makecell{$\displaystyle \frac{\tilde{\rho}_s \tilde{e} \tilde{g} \tilde R^2}{\tilde F_{\rm bv}}$} &  weight/elastohydrodynamic & - \\
$\mathcal I_{\rm bv}$ & \makecell{$\displaystyle \frac{\tilde{\rho}_s \tilde{e} \tilde \omega^2 \tilde R^4}{\tilde{B}}$} & solid inertia/elastohydrodynamic & - \\
$\mathrm{Re}_{\rm bv}$ & \makecell{$\displaystyle \frac{\tilde{\rho}_a \tilde \omega^2 \tilde R^4}{\tilde F_{\rm bv}}$} & fluid inertia/elastohydrodynamic & - \\
$\mathrm{Sq}_{\rm bv}$ & \makecell{$\displaystyle \frac{\tilde F_{\rm bv}}{\tilde{p}_a \tilde R^2}$} & compressibility (elastohydrodynamic/atmospheric pressure) & - \\
$\ell$ & \makecell{$\displaystyle {\tilde \ell}/{\tilde R}$} & forcing size & - 
\end{tabular}
\end{table}

\section{Incompressible and inertialess analysis}
\label{sec:viscous}

\subsection{Weak active forcing ($\alpha \lesssim1 $)}
\label{sec:alphasmall}

\begin{figure}
	 \centering
\includegraphics[width=1.0\textwidth]{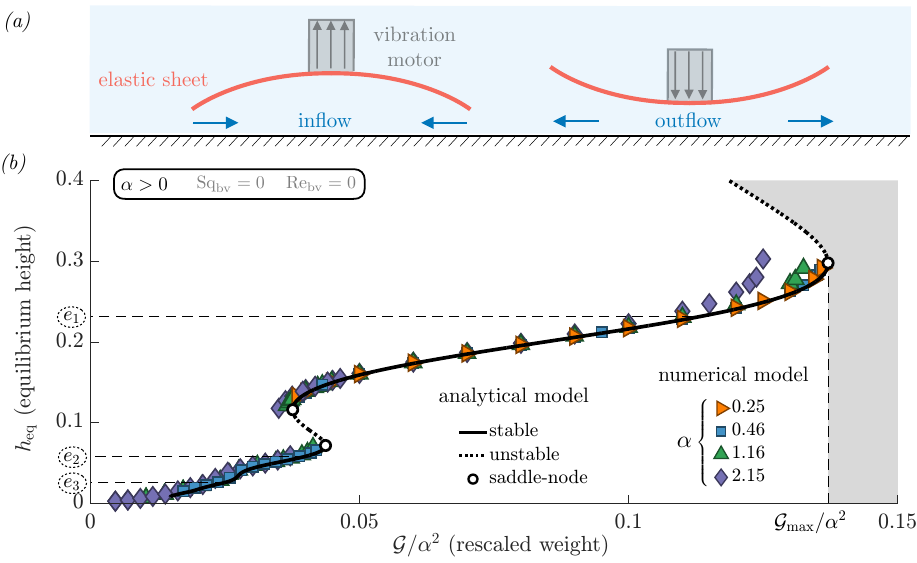}
	\caption{\label{fig:bifurcation_alpha0}
Asymptotic results for $\alpha \lesssim 1$, $\mathcal I_{\rm bv}=\Reynbv=\Sqbv=0$, adapted from \citet{Poulain2024}.
$(a)$ Schematic illustration of the link between the active force direction and the sheet's convexity.
$(b)$ Equilibrium height $\heq$ as a function of the the rescaled dimensionless weight $\mathcal G/\alpha^2$. Symbols are results from numerical simulations, the lines are the prediction of \eqref{eq:gij} obtained by numerical continuation (with a cutoff $N=5$). For $\mathcal G/\alpha^2>\mathcal G_{\rm max}/\alpha^2\simeq 0.137$, no equilibrium is possible and the sheet always detaches from the substrate (grayed area).
}
\end{figure}

We first consider the regime for which inertia and compressibility are negligible, $\mathcal I_{\rm bv}=\Reynbv=\Sqbv=0$, so that the dynamics is solely governed by viscous elastohydrodynamic interactions. Then, the governing equations \eqref{eq:massbalance2} and \eqref{eq:fluxRe} simplify to the Reynolds equation describing inertialess incompressible lubrication flows:
\begin{align}
    12\partiald ht-\gradperp \cdot \left(h^3 \gradperp p\right)=0.
    \label{eq:Reynolds}
\end{align}
We studied this regime in \citep{Poulain2024} in the limit $\alpha \lesssim 1$.
In short, when an elastic sheet is driven by periodic vibrations at its center, it tends to adhere to a nearby surface due to the coupling between its elastic deformation and the lubrication flow in the intervening gap.
As the sheet is pushed towards the surface, it adopts a convex shape that favors fluid outflow; while when it is pulled away from the surface, it adopts a concave shape that resists inflow (\cref{fig:bifurcation_alpha0}$a$). This asymmetry in the flow response over a period of oscillation results in a net inflow and leads to a time-averaged attraction toward the surface.
This mechanism can counteract the sheet’s weight and give rise to an equilibrium hovering height.
We have used these insights to study the system analytically and give the conclusions of our analysis below. We refer the reader to \cref{sec:appendixWeak} for further details on the assumptions we have used, and to \cite{Poulain2024} for the complete derivation.
In short, the height $h(x,t)$ is decomposed into spatial eigenmodes, and an asymptotic analysis allows to find an evolution equation for time-averaged height $\langle h_0 \rangle(t) = \int_t^{t+2\pi}h(x=0,t')~{\rm d}t'$ at $\OO{\alpha^2}$:
\begin{align}
\begin{split}
	\frac{1}{\alpha^2 \langle h_0\rangle ^2} \frac{{\rm d}\langle h_0\rangle }{{\rm d}t}&= 
	\frac{1}{4}\frac{\mathcal G}{\alpha^2} \langle h_0\rangle
	- d_0
	+ \sum_{i,j=1}^{\infty} d_{ij} g_{ij}\left(\langle h_0 \rangle \right),\\
\quad g_{ij}(h)&= \frac{1+\left(\dfrac{h}{\sqrt{e_ie_j}}\right)^6}{\left(1+\left(\dfrac{h}{e_i}\right)^6\right)\left(1+\left(\dfrac{h}{e_j}\right)^6\right)}.
    \label{eq:gij}
\end{split}
\end{align}
The first two terms on the right-hand side of \eqref{eq:gij} correspond respectively to the effect of gravity, where we assumed $\mathcal G/\alpha^2 = \OO1$, and to the first-order effect of elastohydrodynamics, with $d_0=0.0122$.
The third term corresponds to the effects of the eigenmodes $\zeta_i$ (shown in \cref{fig:bifurcation_alpha0_new22}) on the dynamics, with the $e_i=0.242/i^2$,
characteristic dimensionless heights corresponding to the height scale below which the $i$-th mode $\zeta_i$ is excited. This means that we predict higher-order excitation modes as the sheet approaches the wall.
The $d_{ij}$ are numerical coefficients of the symmetric matrix $d_{ij}$ characterizing the strength of the effect of mode $i$ ($i=j$) or of the coupling between modes $i$ and $j$ on the dynamics; their values are given in \cref{sec:appendixWeak}.

\begin{figure}
	 \centering
\includegraphics[width=\textwidth]{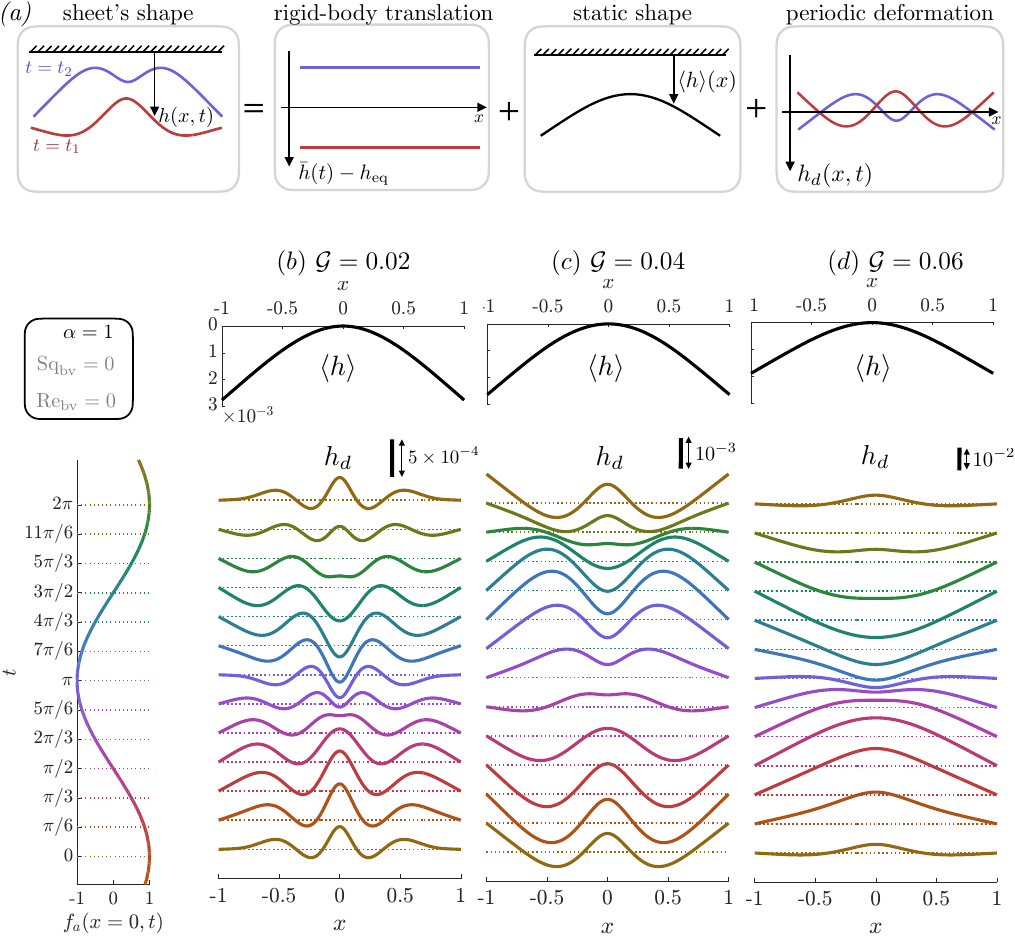}
	\caption{\label{fig:shape_alpha0}
$(a)$ Illustration of the decomposition \eqref{eq:decomposition} of the sheet's shape into a static shape $\langle h \rangle (x)$ (independent of time at the time-averaged steady state), a rigid-body translation $\bar h(t)-h_{\rm eq}$, and a time-periodic deformation $h_d(x,t)$. 
$(b,c,d)$
Time-averaged shape $\langle h \rangle$ and the periodic deformation $h_d$ for $G=0.02,~0.04,~0.06$, respectively, and $\alpha=1$, $\mathcal I_{\rm bv}={\rm Re}_{\rm bv}={\rm Sq}_{\rm bv}=0$.
These are obtained from numerical simulations once a time-averaged steady state has been reached.
$h_d$ is shown at various times of one vibration cycle, with scale bars showing the amplitude of the deformations.
As $\mathcal G$ decreases, higher-order vibration modes are excited.}
\end{figure}

The bifurcation diagram of \eqref{eq:gij} is shown in \cref{fig:bifurcation_alpha0}$(b)$.
If the rescaled weight is too large, $\mathcal G>\mathcal G_{\rm max}=0.137 \alpha^2$, there is no equilibrium and the sheet fails to adhere: gravity overcomes the adhesive elastohydrodynamic effect and $\langle h_0 \rangle \rightarrow \infty$.
For $0<\mathcal G<\mathcal G_{\rm max}$, the weight and the adhesive elastohydrodynamic effect can be in balance and the sheet finds an equilibrium near the wall, $\langle h_0 \rangle \rightarrow h_{\rm eq}$.
We show in \cref{fig:bifurcation_alpha0}$(b)$ that numerical simulations of the governing equations \eqref{eq:normalForceBalance_dimensionless} and \eqref{eq:Reynolds} agree remarkably well with the bifurcation diagram of the reduced model \eqref{eq:gij} for $\alpha \lesssim 1$.

To study the sheet's deformations numerically, we consider the time-averaged equilibrium and let
\begin{align}
    h(x,t)=h_d(x,t)+\langle h\rangle (x,t) + \bar h(t)-h_{\rm eq},
    \label{eq:decomposition}
\end{align}
where $\langle h \rangle (x,t)=(1/2\pi)\int_t^{t+2\pi} h(x,t')~{\rm d}t'$ is the time average (independent of time once a time-averaged steady state is reached), $\bar h (t)=(1/2) \int_{-1}^{1} h(x',t)~{\rm d}x'$ is the spatial average, and $h_{\rm eq}=\langle \bar h \rangle$ the time- and space-averaged height.
The function $h_d(x,t)$ then represent the periodic deformations to the static shape $\langle h \rangle (x)$, with $\langle h_d \rangle = \bar h_d = 0$. The decomposition \eqref{eq:decomposition} is illustrated in \cref{fig:shape_alpha0}$(a)$, where we also show examples of deformations from numerical simulations in panels $(b-d)$.
This illustrates that higher-order modes of deformation are indeed excited as $\mathcal G/\alpha^2$ and $h_{\rm eq}$ decrease. 

Finally, in terms of scaling, our analysis shows that for $\alpha \lesssim 1$ the maximum dimensionless supported weight $\mathcal G_{\rm max}$ scales as $\alpha^2$, the square of the dimensionless forcing strength.
Coming back to dimensional quantities gives the following scaling: $\tilde W_{\rm max} \sim \tilde F_a^2/\tilde F_{\rm bv}$, with $\tilde F_{\rm bv}$ the elastohydrodynamic force scale defined in \eqref{eq:Hbv} and $\tilde F_a$ the active force.
This predicts that, for a given forcing, using an increasingly soft sheet would allow to sustain an arbitrarily large weight regardless of the forcing. However, as the sheet becomes softer ($\tilde F_{\rm bv}$ decreases), $\alpha=\tilde F_a/\tilde F_{\rm bv}$ increases and the assumption $\alpha \lesssim 1$ breaks down.
We study this second regime next.

\subsection{Strong active forcing ($\alpha \geq 1$)}

\label{sec:contactless}
\subsubsection{Contactless adhesion -- Maximum supported weight}

\begin{figure}
	 \centering     \includegraphics[width=0.95\textwidth]{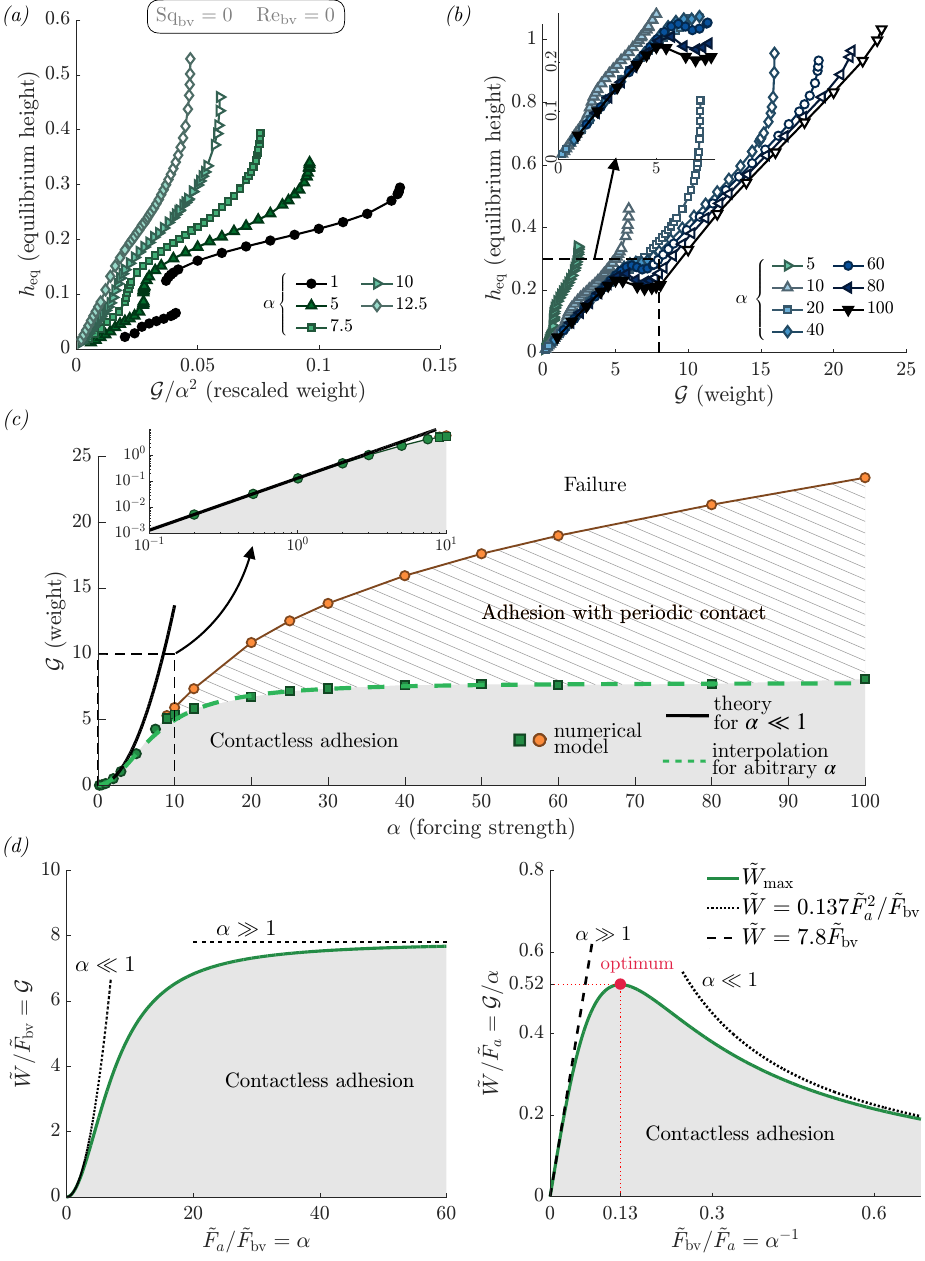}
	\caption{
    Varying active forcing $\alpha$ for an inertialess and incompressible system, $\mathcal I_{\rm bv}={\rm Re}_{\rm bv}={\rm Sq}_{\rm bv}=0$. \label{fig:numerics_alphalarge}
$(a,b)$ Equilibrium height as a function of the dimensionless weight. As $\alpha$ increases, the scaling $\mathcal G_{\rm max} \sim \alpha^2$ does not collapse the data anymore. Open symbols represent cases where contact occurs at the sheet's edges.
$(c)$ Regime map showing the three different possibilities (adhesion with or without edge contact, and adhesion failure) as a function of the dimensionless weight and forcing. The solid line represents the prediction $\mathcal G_{\rm max}=0.137\alpha^2$ derived for $\alpha \ll 1$, and the dashed line is the interpolation \eqref{eq:interpolation}. The inset is a zoom near the origin on a logarithmic scale. (d) Regime maps from \eqref{eq:interpolation} as $\tilde F_a$ varies for fixed $\tilde F_{\rm bv}$ (left), and $\tilde F_{\rm bv}$ varies for fixed $\tilde F_a$ (right).
}
\end{figure}

We show in \cref{fig:numerics_alphalarge}$(a,b)$ bifurcation diagrams obtained by numerically solving the governing equations \eqref{eq:normalForceBalance_dimensionless} and \eqref{eq:Reynolds} for $0<\alpha\leq100$, $\mathcal I_{\rm bv}={\rm Re}_{\rm bv}={\rm Sq}_{\rm bv}=0$.
The corresponding regime map of accessible dimensionless weight is shown in \cref{fig:numerics_alphalarge}$(c)$ and reveals three distinct regions: (i) a regime where adhesion is not possible and the system fails, (ii) a regime of contactless adhesion, and (iii) a regime where we predict adhesion but with the sheet's edges periodically touching the substrate.
The latter effect is discussed in \S \ref{sec:contact}; we first focus on contactless adhesion. For a given $\alpha$, we define $\mathcal G_{\rm max}$ as the threshold weight above which contactless adhesion cannot take place.

For $\alpha \lesssim 1$, the relation $\mathcal G_{\rm max}=0.137 \alpha^2$ is well verified as discussed previously, but becomes a large overestimation of $\mathcal G_{\rm max}$ as $\alpha$ increases.
We see in \cref{fig:numerics_alphalarge}$(c)$ that $\mathcal G_{\rm max}$ saturates for  $\alpha \gtrsim 20$, approaching a constant value $\mathcal G_{\rm max} \simeq 7.8$.
These two asymptotic behaviors are captured by the following interpolation, which also captures well the numerical results over the whole explored range $0<\alpha\leq 100$ (\cref{fig:numerics_alphalarge}$c$):
\begin{align}
    \mathcal G_{\rm max} \simeq 0.137 \frac{\alpha^2}{1+0.0176\alpha^2}, \quad
    \tilde W_{\rm max}\simeq 0.137 \dfrac{\tilde F_a^2/\tilde F_{\rm bv}}{1+0.0176 \tilde F_a^2/\tilde F_{\rm bv}^2}.
    \label{eq:interpolation}
\end{align}
In dimensional quantities, 
we find $\tilde W_{\rm max}=0.137 \tilde F_a^2/\tilde F_{\rm bv}$ for $\tilde F_a \lesssim \tilde F_{\rm bv}$ (weak forcing), and $\tilde W_{\rm max}\simeq 7.8 \tilde F_{\rm bv}$ for $\tilde F_a \gg \tilde F_{\rm bv}$ (strong forcing). 
In other words, if the forcing is strong enough, the maximum supported weight becomes independent of the forcing amplitude and scales as $\tilde F_{\rm bv}$.

In terms of scaling analysis, we expect $\tilde W_{\rm max} \sim \tilde F_a^{2n}/\tilde F_{\rm bv}^{2n-1}$ with $n$ an integer: this involves all the force scales in the system, has the correct dimensions, and recovers the fact that the averaged dynamics of the system should only depend on even powers of $\tilde F_a$, given that a change of sign of $\tilde F_a$ is equivalent to a change of phase of the forcing which cannot affect the long-term dynamics.
The asymptotic regime for $\alpha \lesssim 1$ corresponds to $n=1$: increasing the magnitude of the active force first increases the maximum weight as $\tilde W \sim \tilde F_a^2/\tilde F_{\rm bv}$.
However, as $\tilde F_a$ reaches $\tilde F_a \sim \tilde F_{\rm bv}$, the maximum weight eventually saturates: $n=0$, $\tilde W\sim \tilde F_{\rm bv}$ for $\alpha \gg 1$.

\Cref{eq:interpolation} provides useful insights into the evolution of the maximum supported weight $\tilde W_{\rm max}$ as a function of the forcing amplitude $\tilde F_a$, for fixed $\tilde F_{\rm bv}=(\tilde \mu \tilde \omega \tilde B^2)^{1/3}$, or conversely as a function of $\tilde F_{\rm bv}$ for fixed $\tilde F_a$, assuming both forces can vary independently.
This is shown in \cref{fig:numerics_alphalarge}(d).
At fixed $\tilde F_{\rm bv}$, increasing $\tilde F_a$ initially increases significantly the maximum weight, following $\tilde W_{\rm max}\sim \tilde F_a^2$. However, beyond $\tilde F_a/\tilde F_{\rm bv}=\OO{10}$, $\tilde W_{\rm max}$ saturates and becomes insensitive to the forcing.
Conversely, at fixed $\tilde F_a$, decreasing $\tilde F_{\rm bv}$ (e.g., by making the sheet softer) initially increases the maximum weight as $\tilde W_{\rm max} \sim \tilde F_{\rm bv}^{-1}$.
Beyond a threshold, however, the sheet becomes too soft, and making it even softer decreases the maximum weight as $\tilde W_{\rm max}\sim \tilde F_{\rm bv}$. This shows an optimal force $\tilde F_{\rm bv}^\star \approx 0.13 \tilde F_a$ that maximizes the maximum supported weight, and an associated optimal bending stiffness $\tilde B^\star \approx 0.05 \tilde F_a^{3/2} \left(\mu \omega\right)^{-1/2}$. The same procedure could be followed to derive an optimal frequency $\tilde \omega$, however, changing forcing frequency would usually also alter the forcing amplitude $\tilde F_a$ so that the optimal frequency would, in practice, depend on the details of the actuation. For instance, Linear Resonant Actuators (LRA) are a type of vibration motor that typically only work at a fixed frequency, while Eccentric Rotating Mass (ERM) motors are such that $\tilde F_a \sim \tilde \omega^2$.

If both $\tilde F_{\rm bv}$ and $\tilde F_a$ can be tuned independently, the optimal $\tilde W_{\rm max}$ is achieved by maximizing $\tilde F_a$.
In practice, this could be achieved by selecting the most powerful available periodic actuator, which sets both $\tilde F_a$ and $\tilde \omega$.
The sheet's bending rigidity $\tilde B$ can then be adjusted, by varying its thickness or Young's modulus, to ensure it is equal to $\tilde B^\star$.
Practical design would also need to account for energy consumption and for the fact that powerful vibration motors are typically both heavy and bulky. Their lateral extent, neglected in this study, also limits $\tilde W_{\rm max}$ \citep{Poulain2024}.
Nevertheless, our prediction resonates qualitatively with experimental observations: using relatively small motors, \cite{Weston2021} demonstrated  $\tilde W_{\rm max}\approx 5~\unit{\newton}$ using thin and soft sheets, while experiments by \citet{ColasanteYT} (see also experiments from the same author reported by \citet{Ramanarayanan2024}) show $W_{\rm max}=\OO{100\unit{\newton}}$ using much larger designs, stronger motors, and stiffer sheets.

\subsubsection{Contactless adhesion -- Equilibrium height}
We observe a remarkable linear relationship between $ h_{{\rm eq}}$ and $\mathcal G$ for $\alpha \gg 1$ in the inset of \cref{fig:numerics_alphalarge}$(b)$, for $\mathcal G \lesssim \mathcal G_{\rm max}$.
Combining this observation with the asymptotic results of the previous section, we find:
\begin{subequations}
\begin{align}
     h_{\rm eq} \approx
    \begin{cases}
        f \left(\dfrac{\mathcal G}{\alpha^2}\right)~~~~\text{if}~~~~ \alpha \lesssim 1, \\[20pt]
        0.05 \mathcal G ~~~~\text{if}~~~~ \alpha \gg 1,
    \end{cases}
    ~~\text{for}~\mathcal G\lesssim\mathcal G_{\rm max}.
\end{align}
or, in dimensional units,
\begin{align}
    \tilde h_{\rm eq} \approx
    \begin{cases}
        \tilde H_{\rm bv} \times f \left(\dfrac{\tilde W \tilde F_{\rm bv}}{\tilde F_a^2}\right)~~~~\text{if}~~~~ \tilde F_a \lesssim \tilde F_{\rm bv}, \\[20pt]
        0.05 \dfrac{\tilde W \tilde R^2}{\tilde B } ~~~~\text{if}~~~~ \tilde F_a \gg \tilde F_{\rm bv}
    \end{cases}
    ~~\text{for}~\tilde W\lesssim\tilde W_{\rm max},
    \label{eq:scalingHeight}
\end{align}
\end{subequations}
with the maximum weight given by \eqref{eq:interpolation}.
The function $f$ represents the stable fixed points of \eqref{eq:gij} and is shown in \cref{fig:bifurcation_alpha0}$(a)$; its analytical approximations are discussed by \citet{Poulain2024}.
\Cref{eq:scalingHeight} shows that, for strong forcing $\alpha \gg 1$, the equilibrium height results from a balance between the bending force and gravity, with viscosity setting the maximum supported weight.
Remarkably, the equilibrium height at the maximum supported weight is  $\tilde h_{\rm eq}(\tilde W_{\rm max}) \simeq 0.3 \tilde H_{\rm bv}$ for $\alpha \gg 1$, which is also true with a similar prefactor for $\alpha \ll 1$ (\cref{fig:bifurcation_alpha0}$b$), so that the following holds over the range $0<\alpha \leq 100$:
\begin{align}
    \tilde h_{\rm eq}(\tilde W_{\rm max}) \approx 0.3 \tilde H_{\rm bv}.
\end{align}
While it is not straightforward to rationalize the linear relation between $h_{\rm eq}$ and $\mathcal G$, we note that $\tilde B/\tilde R^2 = \tilde F_{{\rm bv}}/\tilde H_{\rm bv}=\tilde k_{\rm bv}$, with $\tilde k_{\rm bv}$ a natural scale for stiffness, analogous to a spring constant. Then,  \eqref{eq:scalingHeight} for $\alpha \gg 1$ can be written $\tilde h_{\rm eq} \approx 0.05 \tilde W/\tilde k_{\rm bv}$.

\begin{figure}
	 \centering
     \includegraphics[width=\textwidth]{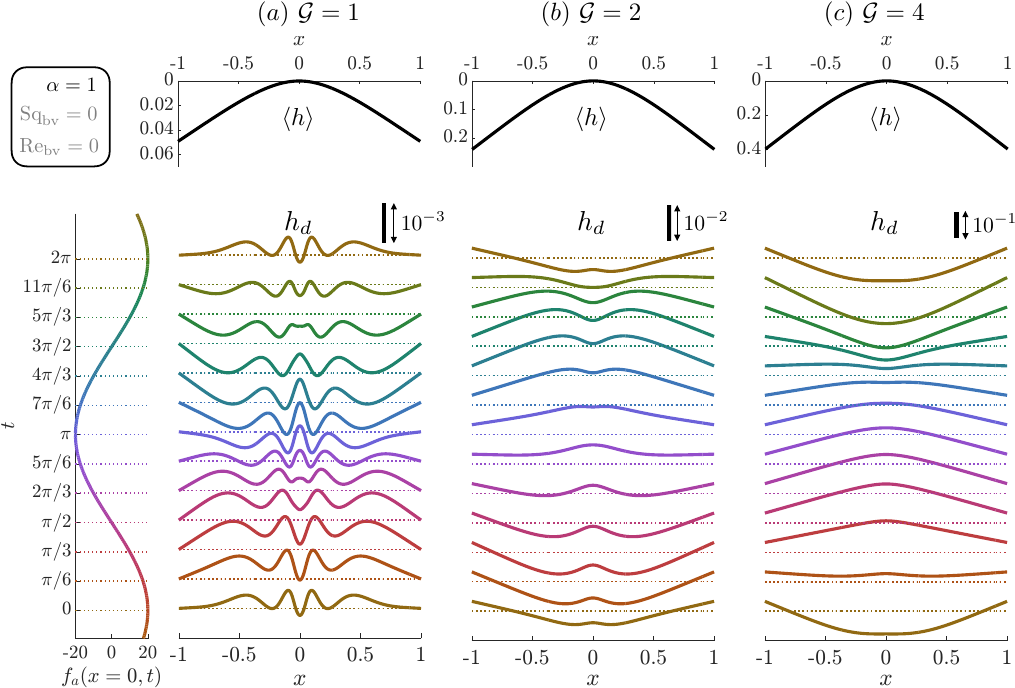}
	\caption{\label{fig:alpha20}
Time-averaged shape $\langle h \rangle$ and the periodic deformation $h_d$ for $\alpha=20$, $\mathcal I_{\rm bv}=\Reynbv=\Sqbv=0$, and $\mathcal G=1,~2,~4$ in $(a,b,c)$, respectively.}
\end{figure}

\subsubsection{Contactless adhesion -- Sheet's deformations}

\begin{figure}
	 \centering
     \includegraphics[width=0.8\textwidth]{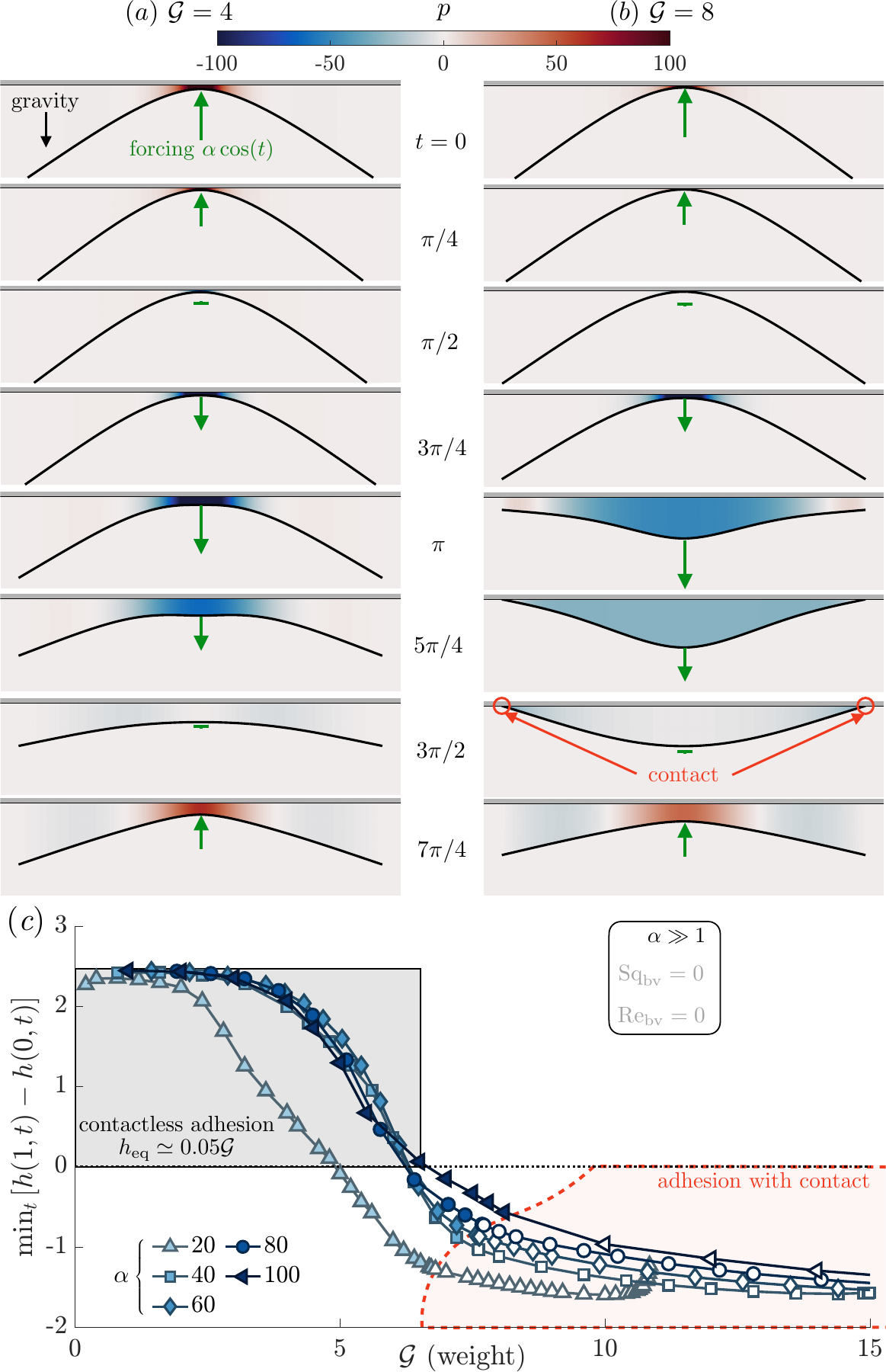}
	\caption{\label{fig:touching}
$(a,b)$ Sheet's shape and pressure field over a period of vibration for $\alpha=20$ and $(a)$ $\mathcal G=4$, $(b)$ $\mathcal G=8$.
The green arrows represent the active force schematically, periodically pushing and pulling at the center of the sheet. In $(b)$, at $t=3\pi/2$, the edges of the sheet touch the bottom wall.
$(c)$ The difference in height between the sheet's center $h(0,t)$ and its edge $h(1,t)$ is a measure of the sheet's convexity: when ${\rm min}_t(h(x=1,t)-h(x=0,t))>0$, the sheet always remain convex as in panel $(a)$, and the linear relationship $h_{\rm eq}\simeq 0.05 \mathcal G$ between equilibrium height and weight is verified for $\alpha \gg 1$ (see \cref{fig:numerics_alphalarge}$b$). The existence of a concave part during the vibration cycle, as shown at $t=\pi,~5\pi/4$ and $3\pi/2$ in panel $(b)$, is associated with contact with the wall for $\alpha \gg 1$. Filled symbols represent the case when the sheet never touches the wall, open symbols correspond to the sheet periodically touching the wall.
}
\end{figure}

Some of the insights gained from the weak forcing analysis presented in \S \ref{sec:alphasmall} carry over to the regime $\alpha \gg 1$. In particular, we show in \cref{fig:alpha20} that the excitation of higher-order modes of deformation as $\mathcal G\rightarrow 0$, $h_{\rm eq}\rightarrow 0$ remains valid.
One important difference, however, is the amplitude of the sheet's deformation.
In particular, we show in \cref{fig:touching}$(a,b)$ an illustration of the sheet's shape for $\alpha=20$. For $\mathcal G<\mathcal G_{\rm max}$, panel $(a)$, the sheet remains convex, as expected from the large gravitational pull alone. For $\mathcal G>\mathcal G_{\rm max}$, panel $(b)$, the sheet becomes concave when the active forcing pulls it away from the wall. This, in turn, causes the sheet's edge to come into contact with the substrate.
We confirm in \cref{fig:touching}$(c)$ that turning the sheet concave during part of the vibration cycle is what leads to contact with the wall, and therefore what sets $G_{\rm max}$.
This change in convexity occurs when the active pull is maintained for a sufficiently long time compared to the time it takes for the sheet to undergo a significant change in shape. This process is associated with a timescale $\tilde t_v$, which can be estimated in terms of scaling from  \eqref{eq:Reynolds}:  $\mu \tilde h_{\rm eq}/\tilde t_v\sim \tilde p_g \tilde h_{\rm eq}^3/\tilde \mu \tilde R^2$, with $\tilde h_{\rm eq}\sim \tilde W/\tilde k_{bv}$ from \eqref{eq:scalingHeight}  and the pressure due to gravity $\tilde p_g\sim\tilde W/\tilde R^2$. This gives $\tilde t_v \sim \tilde \mu \tilde B^2 /\tilde W^3$.
The criterion for contact is then   $\tilde t_v \tilde \omega \sim (\tilde F_{\rm bv}/\tilde W)^3 \lesssim 1$, and we recover the scaling \eqref{eq:interpolation} for the maximum supported weight for $\alpha \gg 1$, i.e., $\tilde W_{\rm max} \sim \tilde F_{\rm bv}$.
We note that this understanding relies heavily on the convexity of the sheet and, consequently, on the specific distribution of weight and of active force. While we have assumed a uniform weight distribution, the effects of non-uniform weight remain to be explored in future work.

\subsubsection{Adhesion with contact}
\label{sec:contact}
For $\alpha\gtrsim 9$ and sufficiently large $\mathcal G$, we observe in \cref{fig:numerics_alphalarge}$(b,c)$ that adhesion can be maintained while the sheet's edges come in periodic contact with the substrate. 
The dynamics in this regime is illustrated in \cref{fig:touching}.
To handle contact numerically, we have introduced the repulsive potential $f_w(h)$ in \eqref{eq:normalForceBalance_dimensionless}, which models an elastic collision.
Although numerical simulations can access the contact regime, contact introduces additional physical effects beyond the scope of the present model. This regime is therefore not examined further in this article.
Nonetheless, we want to point to a possible analogy with suction cups \citep{ramanarayanan2024emergence}, which achieve strong adhesion through edge contact and pressure differentials. Suction cups are prone to failure, with small perturbations at its edges leading to detachment \citep{Tiwari2019}.
It has been proposed that applying vibrations to suction cups could improve their performance and reliability \citep{Zhu2006,Hong2009}, an idea that shares similarities with the contact regime of vibrated sheets described above.
An experimental confirmation of this regime would be valuable and could guide further investigations into its modeling and underlying dynamics.

\section{Inertial effects: influence of the Reynolds number}
\label{sec:inertia}

We now introduce the effects of a finite Reynolds number $\Reynbv=\tilde \rho_a \tilde \omega \tilde R^4/\tilde B>0$ as defined in \eqref{eq:dim1} in the dynamics of adhesion, for an incompressible fluid ($\Sqbv=0$) and an inertialess sheet ($\mathcal I_{\rm bv}=0$).
The governing equations are then \eqref{eq:massbalance2} with $\rho=1$, \eqref{eq:fluxRe} which corrects the lubrication equations to account for fluid inertia as discussed previously, and the force balance \eqref{eq:normalForceBalance_dimensionless}.
Using the parameter values described in \S \ref{sec:governing}, $\Reynbv = \OO{10^4}$. We will see that the actual effects of fluid inertia are, in fact, more properly characterized by $\Reyneq=h_{\rm eq}^2 \Reynbv \lesssim \OO{10}$, the Reynolds number that uses the equilibrium height as the characteristic length.
Further, $\Reynbv$ depends strongly on the details of the experiment (vibration frequency, sheet's radius and bending modulus), and both the regimes of small and large $\Reynbv$ can be relevant depending on the system's scale.

\subsection{Rigid sheets}
\label{sec:inertiaRigidSheet}
\begin{figure}
	 \centering
     \includegraphics[width=1.0\textwidth]{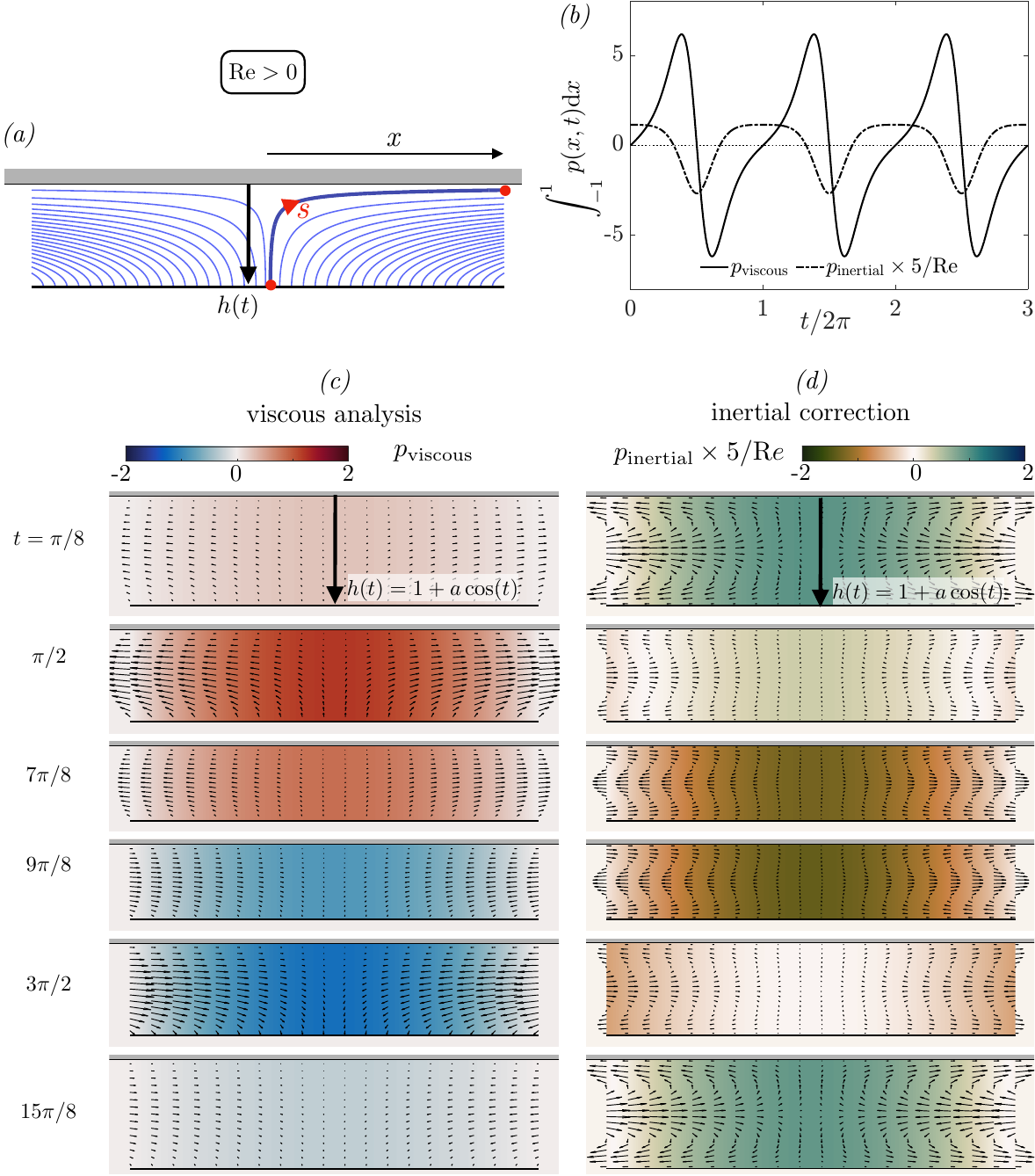}
	\caption{
    Squeeze flow of a rigid plate moving normal to a wall with a height evolving as $h(t)=1+a\cos(t)$, shown for illustration here with $a=0.4$. The pressure is computed from \eqref{eq:pressureMainText} and the fluid velocity from the calculations carried out in \cref{sec:derivationNS}.
    $(a)$ Streamlines of the flow associated with a rigid sheet moving towards a wall or away from it.
    $(b)$ Integral of pressure in space for the viscous component and inertial component.  Integrated in time, the viscous component averages to zero while the inertial component gives a positive force. The inertial pressure is rescaled by the Reynolds number.
    $(c,d)$ Velocity profiles (arrows) and pressure field (colors) isolating $(c)$ the dominant viscous flow and $(d)$ the inertial corrections.
	\label{fig:ReynoldsIntuition}
	}
\end{figure}

To qualitatively understand the effect of fluid inertia, we begin with an idealized, simplified setup: a standard squeeze film where the gap between the wall and a rigid sheet oscillates harmonically. We consider the flow in the gap and project the Navier-Stokes momentum equations along a streamline:
\begin{align}
	 \tilde \rho\left(\partiald{ \tilde v_s}{\tilde t}+ \tilde v_s\partiald{ \tilde v_s}{ \tilde s}\right)=-\partiald{ \tilde p}{ \tilde s} +  \tilde \mu  \tilde \nabla^2  \tilde v_s,
\end{align}
where $\tilde s$ and $\tilde v_s$ refer to the position and velocity along the streamline, respectively.
We integrate this equation along a streamline whose ends are at the top of the sheet, close to $x=0$, and within the gap at a position $\tilde s$, as shown in figure \cref{fig:ReynoldsIntuition}$(a)$, to find an energy balance analogous to Bernoulli's principle:
\begin{align}
 \left( \tilde p\big\rvert_{\ddim s=0}+\frac{1}{2} \tilde \rho  \tilde v_s^2\big\rvert_{\ddim s=0}\right) - \left( \tilde p\big\rvert_{\ddim s}+\frac{1}{2} \tilde \rho \tilde v_s^2\big\rvert_{\ddim s}\right)+\int_0^{\tilde s} \tilde \rho \partiald{ \tilde v_s}{ \tilde t}\biggr\rvert_{\ddim s'}{\rm d} \tilde s'=\Delta  \tilde p_{\rm visc},
 \label{eq:bernou}
\end{align}	
with $\Delta  \tilde p_{\rm visc}=\int_0^{\tilde s}  \tilde \mu \tilde \nabla^2  \tilde v_s~{\rm d} \tilde s'$ the pressure loss due to viscous stresses. 
 According to lubrication scalings: $ \tilde v_s\big\rvert_{\ddim s=0} = \tilde{\dot{h}} \ll \tilde v_s\big\rvert_{\ddim s}\simeq - \tilde{\dot{h}}\tilde x/\tilde h$, with overdots denoting time derivatives.
 We use $\tilde s \simeq \tilde x$ away from $\tilde s=0$ since the flow in the gap is predominantly horizontal under the lubrication approximation.
Then, in terms of scaling, $\int_0^{\tilde s} (\partial{ \tilde v_s}/\partial{ \tilde t}){\rm d} \tilde s' \sim [(\tilde{\dot h})^2-\tilde{\ddot h}/\tilde h] \, \tilde x^2$ and
$\Delta  \tilde p_{\rm visc}  \sim -\tilde \mu \tilde{\dot h}\tilde x^2/\tilde h^3$.
\Cref{eq:bernou} together with these scaling, gives the following pressure profile: $\tilde p(\tilde x,\tilde t) \sim (\ddim R^2-\ddim x^2)[-\tilde\mu \ddim{\dot h}/\ddim h^3 - \ddim \rho {\ddim{\ddot h}}/{\ddim h}+\ddim \rho{\ddim{\dot h}^2}/{\ddim h^2} ]$, with prefactors missing in front of each term representing, respectively, viscous effects, unsteady inertia, and convective inertia.
The formal inertial lubrication analysis  from \eqref{eq:fluxRe} leads to the following, in dimensionless form:
\begin{align}
    p(x,t)&=\left(1-x^2\right)\left(-6 \frac{\dot h}{h^3} - \frac{3{\rm Re}}{5} \frac{\ddot h}{h} + \frac{51{\rm Re}}{35}\frac{\dot h^2}{h^2}\right) + p(x=1,t),
    \label{eq:pressureMainText}
\end{align}
with $p(x=1,t)$ the pressure at the edge found from the boundary condition \eqref{eq:bcPressureMain}, ${\rm Re} = \tilde \rho_a \tilde \omega \tilde h_0^2/\tilde \mu $, and $\tilde h_0$ the typical oscillation height.
This is the dimensionless version of the scaling analysis presented above, including prefactors.
The first term describes the viscous losses and is captured by standard lubrication theory. 
The unsteady acceleration term, proportional to $-\ddot h/h$, is equivalent to an added mass effect as discussed below.
The edge pressure tends to decrease the pressure in the gap; however, the dominant effect of inertia is to convert the kinetic energy density of the fluid drawn into (or ejected from) the gap into pressure. 
This is illustrated in \cref{fig:ReynoldsIntuition} where a squeeze flow is imposed by setting the height as $h(t)=1+a\cos(t)$, $0<a<1$.
We show both the pressure profile and velocity vectors in panels $(c)$ and $(d)$ for the purely viscous analysis and the inertial corrections, respectively.
Over one period of oscillations, the net normal force from viscous effects cancels out, as expected: $\int_{-1}^1 \langle p_{\rm viscous}(x,t) \rangle ~{\rm d}x=0$, with $p_{\rm viscous}(x,t)=-6\left(1-x^2\right)\dot h(t)/h^3(t)$.
However, the force from inertial effects is positive, $ \int_{-1}^1 \langle p_{\rm inertial}(x)\rangle ~{\rm d}x = (1-7k/24)a^2{\rm Re}/1.5 + \OO{a^4}>0 $, with $p_{\rm inertial}=p-p_{\rm viscous}$ and $k=0.5$ a loss coefficient introduced in the boundary condition \eqref{eq:bcPressureMain}, showing that inertia leads to a normal force that pushes the sheet away from the wall. This effect has long been known in the context of bearings (e.g., \citet{Kuzma1968,Tichy1970,Jones1975}). 
In addition to modifying the pressure, a non-zero Reynolds number also alters the parabolic velocity profile predicted by the purely viscous theory: the horizontal velocity profile becomes a $6^{\mathrm{th}}$-order polynomial which allows for secondary flows (\cref{fig:ReynoldsIntuition}$c$), a common feature of pulsatile flows \citep{Womersley1955}.
We finally note that the unsteady acceleration term, proportional to $\ddot{h}$, appears as a height-dependent added mass. Indeed, integrating the pressure over the sheet's length $2\tilde R$ gives rise to a force per unit length $-\tilde m_a \tilde{\ddot{h}}$, with $\tilde m_a=4\tilde \rho \tilde R^3/5\tilde h$ that can be interpreted as a added mass per unit length (see also \eqref{eq:reynoldsODE}). In contrast, the added mass of a plate far from any boundary is proportional to $\tilde \rho \tilde R^2$ \citep{Brennen1982}.


While we initially considered a sheet constrained in position for simplicity, we are more interested in the case of a weightless rigid sheet driven periodically by a dimensionless force $\alpha \cos(t)$.
In this case, a two-timescale asymptotic analysis can be used to find the evolution of the height $h(t)$ in the asymptotic limit $\alpha^2{\rm Re} \ll 1$.
We present the derivation in \cref{sec:inertia_asymp}, where we show that the sheet slowly moves away from the wall to accommodate the increase in pressure due to inertia while maintaining a force balance.
The sheet's time-averaged position follows
\begin{align}
    \frac{1}{\alpha^2\langle h \rangle ^2}\ddt{\langle h \rangle}=\frac{1-7k/16}{224}{\rm Re} \langle h \rangle^5,
\end{align}
which indeed shows repulsion from the wall, ${\rm d}\langle h \rangle/{\rm d}t>0$.
We expect that the destabilizing influence of fluid inertia observed in rigid systems qualitatively extends to soft sheets, weakening the viscous adhesion mechanism discussed in the previous section.

\begin{figure}
	 \centering
     \includegraphics[width=\textwidth]{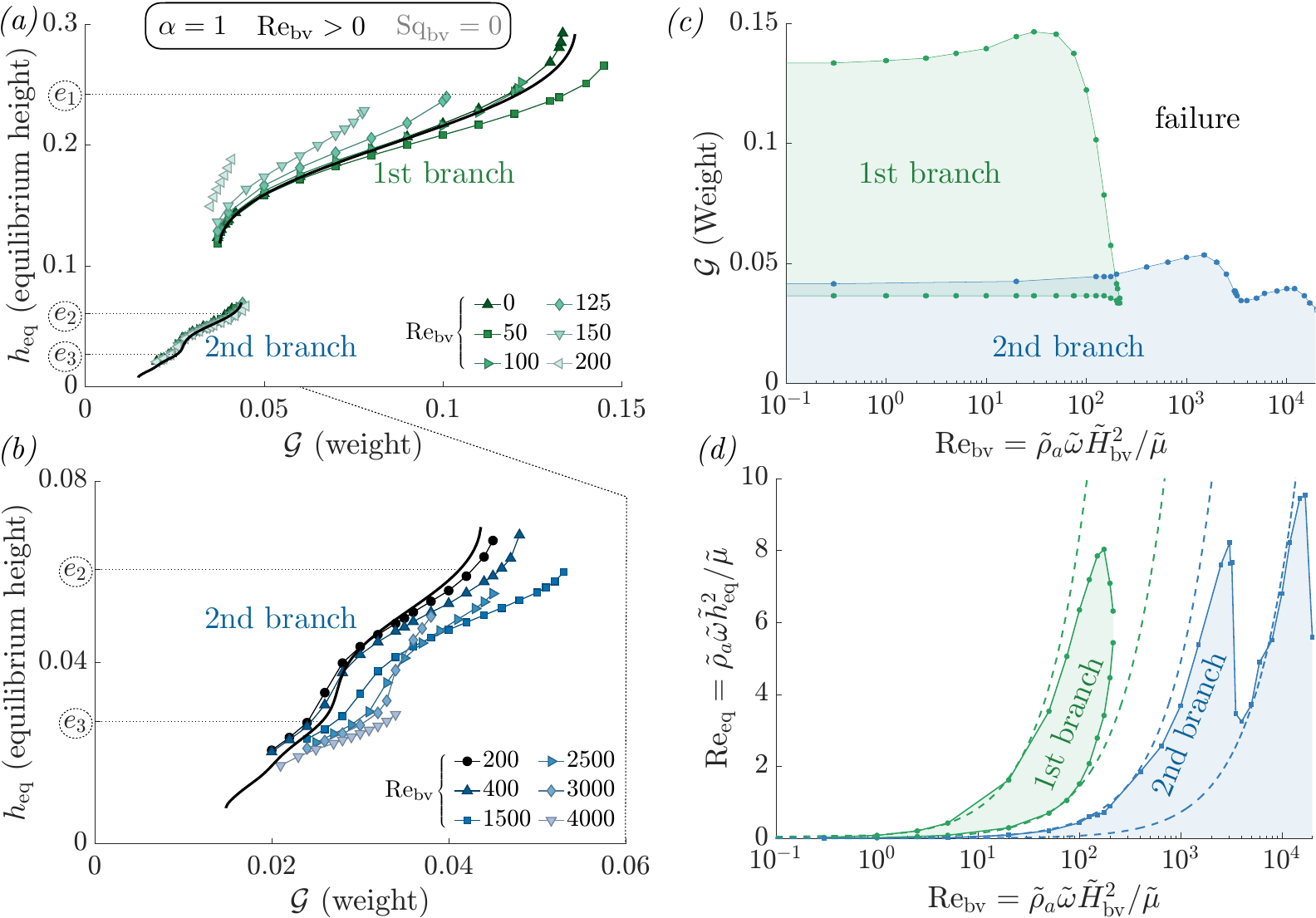}
	\caption{\label{fig:ReynoldsSoft_alpha1}
$(a-b)$ Equilibrium height as a function of the dimensionless weight with $\Sqbv=\mathcal I_{\rm bv}=0$ and $\alpha=1$ for $(a)$ $\Reynbv<200$ and $(b)$ $\Reynbv>200$. 
Black lines are the stable equilibria of \eqref{eq:gij}.
$(c)$ Phase diagram showing the accessible weights as a function of $\Reynbv$. The first equilibrium branch corresponds to $h_{\rm eq}>0.1$, the second branch to $h_{\rm eq}<0.1$.
$(d)$ Reynolds number based on the equilibrium height $h_{\rm eq}$, ${\rm Re}_{\rm eq}=h_{\rm eq}^2 \Reynbv$ as a function of the control parameter $\Reynbv$. The dashed lines represent cases where $h_{\rm eq}$ is constant and illustrate the expected behavior if fluid inertia had no effect on the system. }
\end{figure}

\begin{figure}
	 \centering
     \includegraphics[width=\textwidth]{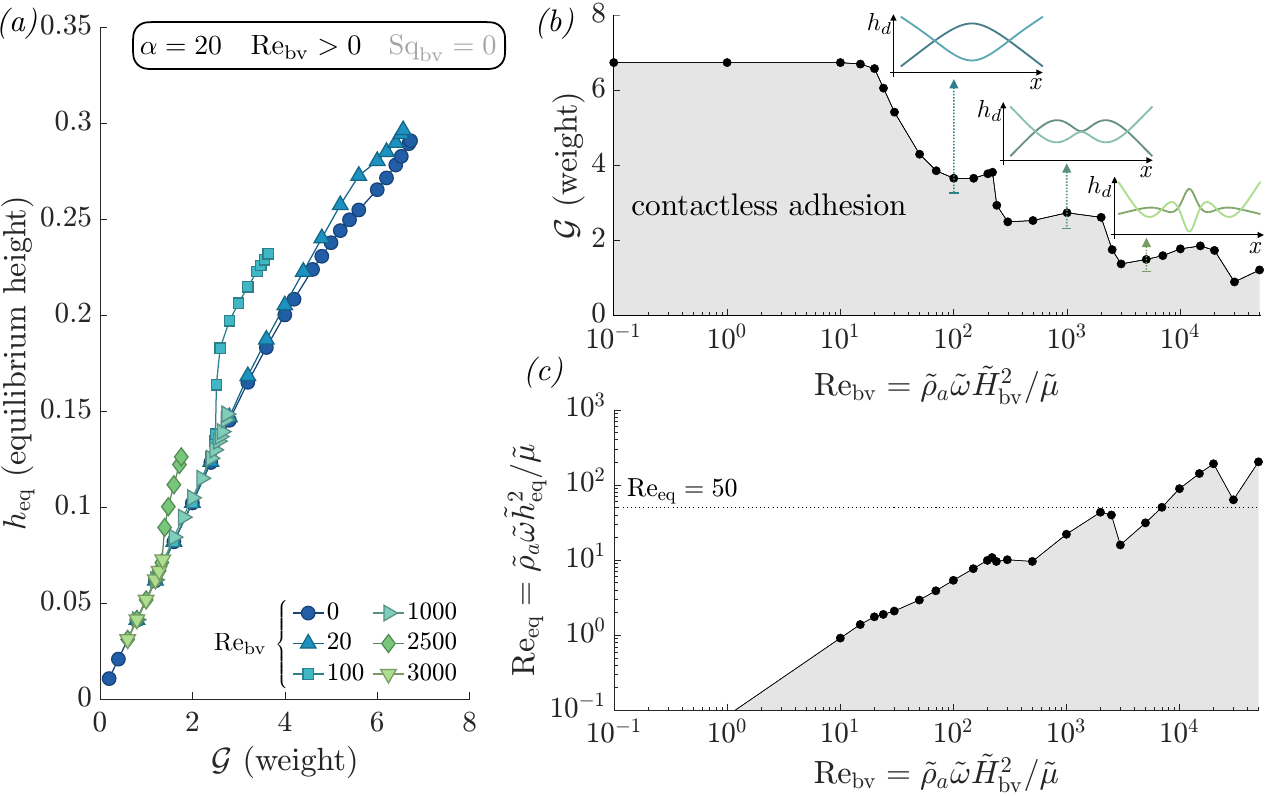}
	\caption{\label{fig:ReynoldsSoft_alpha20}
Effect of the fluid inertia with ${\rm Sq}_{\rm bv}=\mathcal I_{\rm bv}=0$ and ${\rm Re}_{\rm bv}>0$ for $\alpha=20$.
$(a)$ Equilibrium height as a function of the dimensionless weight for the regime of contactless adhesion. 
$(b)$ Regime maps and the associated $(c)$ range of Reynolds number based on equilibrium height $\Reyneq$. The inertial lubrication theory is not expected to be valid for ${\rm Re}_{\rm eq}\gtrsim 50.$
In $(b)$, we show illustrations of $h_d$ (as defined in \eqref{eq:decomposition}) for ${\rm Re}_{\rm bv}$=10, 1000 and 5000 at $t=\pi/2$ and $3\pi/2$.
}
\end{figure}

\subsection{Soft sheets}
We solve the governing equations for soft sheets with $\Sqbv=I_{\rm bv}=0$ and two values of $\alpha$, $\alpha=1$ and $\alpha=20$, corresponding to the regimes of weak and strong forcing, respectively.
For both values, we systematically studied the relationship between the equilibrium height $h_{\rm eq}$ and the weight $G$, as well as the maximum supported weight $ G_{\rm max}$, as a function of the Reynolds number $\Reynbv$.
As we discussed in \S \ref{sec:equili}, for each set of parameters we define the equilibrium Reynolds number ${\rm Re}_{\rm eq}=h_{\rm eq}^2 {\rm Re}_{\rm bv}$ that is based on the actual equilibrium height rather than on the heightscale $\tilde H_{\rm bv}$ and which is representative of the actual magnitude of inertial over viscous effects.

\Cref{fig:ReynoldsSoft_alpha1} summarizes our results for $\alpha=1$.
There, we show examples of bifurcation diagrams and the associated regime map of reachable weights as a function of $\Reynbv$.
For small values of ${\rm Re}_{\rm bv}$, increasing the Reynolds number first marginally increases the maximum weight $\mathcal G_{\rm max}$, up to $\mathcal G_{\rm max}\simeq 0.15$ for ${\rm Re}_{\rm bv} \simeq 40$, to be compared with $\mathcal G_{\rm max}\simeq 0.14$ for ${\rm Re}_{\rm bv} = 0$.
Further increasing ${\rm Re}_{\rm bv}$ then leads to a sharp decrease in $\mathcal G_{\rm max}$, explained by the destabilizing effect of fluid inertia discussed previously.
\Cref{fig:ReynoldsSoft_alpha1}$(d)$ shows the range of accessible equilibrium Reynolds numbers $\Reyneq$ as a function of $\Reynbv$. 
This demonstrates that $\Reynbv$ has little effect on the dynamics as long as $\Reyneq \lesssim 1$. Beyond this, once $\Reyneq=\OO{1}$, $\mathcal G_{\rm max}$ decreases together with $h_{\rm eq}$ to ensure $\Reyneq$ remains relatively small.
In particular, we observe a sharp drop in $\mathcal G_{\rm max}$ for $\Reynbv \gtrsim 30 \approx 1/e_1^2$.
As $\Reynbv$ keeps increasing, the first branch of the bifurcation diagram (continuous black line for ${h_{\rm eq}}>0.1$ in \cref{fig:ReynoldsSoft_alpha1}$a$) is eventually not accessible anymore for ${\rm Re}_{\rm bv} \gtrsim 200 \approx 1/e_2^2$.
We observe a similar drop of $\mathcal G_{\rm max}$ for ${\rm Re}_{\rm bv} \approx 2000 \approx 1/e_3^2$, as the first part of the second equilibrium branch (corresponding to the excitation of the second eigenmode $\zeta_2$ for $h_{\rm eq} \approx e_2$) is also no longer accessible.

The system's behavior is qualitatively similar in the regime of strong forcing, as illustrated in \cref{fig:ReynoldsSoft_alpha20} for $\alpha=20$: the maximum supported weight decreases as ${\rm Re}_{\rm bv}$ increases, with sharp drops at specific Reynolds numbers, while the equilibrium height shows minor variations with ${\rm Re}_{\rm bv}$.
Similarly to the weak forcing case, these drops are associated with higher-order modes as illustrated in panel $(b)$. 
An important difference between the weak and strong forcing regimes is that the equilibrium Reynolds number ${\rm Re}_{\rm eq}$ can become relatively large in the latter, and we observe values up to ${\rm Re}_{\rm eq}=\OO{10^2}$ for $\Reynbv=\OO{10^4}$.
For such large values, the first-order inertial corrections to lubrication theory that we use (\cref{eq:fluxRe} and \cref{sec:derivationNS}) may no longer be valid, and higher-order corrections, or full Navier-Stokes simulations, may be necessary to accurately describe the flow and $\mathcal G_{\rm max}$.
We also make an anecdotal observation that for $200<{\mathrm Re}_{\mathrm bv}<300$ the sheet can respond at half the forcing frequency without altering the adhesion strength. Such period doubling is a well-known feature of nonlinear forced systems, and has been observed and predicted in inertial lubrication flows with free surfaces \citep{Rojas2010}.

\section{Compressible effects: influence of the Squeeze number}
\label{sec:compressible}
In this section, we neglect inertia ($\mathcal I_{\rm bv}={\rm Re}_{\rm bv}=0$) and study the influence of the fluid's compressibility, ${\rm Sq}_{\rm bv}>0$, assuming that it behaves as an isothermal ideal gas.
We assume that the two bounding surfaces, the rigid wall and the elastic sheet, are isothermal, and the characteristic timescale for temperature variations within the thickness of the gap is then $H_{\rm bv}^2/\ddim D$ with $\ddim D$ the thermal diffusivity of the gas.
Comparing this time with the characteristic forcing time $\ddim \omega^{-1}$ defines a Péclet number ${\rm Pe}=\ddim \omega H_{\rm bv}^2/\ddim D_{\rm th}=\Reynbv {\rm Pr}$ with ${\rm Pr}=\ddim \mu/\ddim \rho_a \ddim D_{\rm th}\simeq 0.7$ the Prandtl number.
Our assumption of isothermal gas is appropriate for small ${\rm Pe}$ and therefore for small Reynolds numbers. We make this assumption and therefore use the isothermal ideal gas law \eqref{eq:perfectgas}.
In the opposite limit of an isentropic gas, valid for large Péclet numbers, the pressure-density relation would differ. However, the essential mechanism of the increase in density with increasing pressure would still be valid, such that the results discussed next would be qualitatively similar.

\subsection{Rigid sheets}
To build intuition, we first consider the squeeze film setup with a rigid sheet and have $\tilde h=\tilde h_0(1+a\cos(\tilde \omega \tilde t))$.
As discussed in \S\ref{sec:governing}, the effects of compressibility are quantified with the Squeeze number ${\rm Sq}=\tilde\mu\tilde\omega\tilde R^2/\tilde h_0^2\tilde p_a$, which compares the viscous stresses to the ambient pressure $\tilde p_a$.
When the sheet approaches the wall, a competition arises between the outflow of fluid from the gap, resisted by viscosity, and the compression of the fluid, resisted by its bulk modulus.
For an isothermal ideal gas, the bulk modulus is the pressure.
For ${\rm Sq} \ll 1$, the resistance to compression is much weaker than the resistance to flow, and an incompressible description is appropriate.
Conversely, for ${\rm Sq} \gg 1$, viscous stresses become so large that there is almost no outflow, and the fluid is compressed similarly to a piston.
Likewise, fluid is expanded when the surface moves away from the wall.

In this limit ${\rm Sq} \gg 1$, the work per unit area to expand from $\tilde h_0$ to $\tilde h_0(1+a)$ is $\tilde w_{+a}=\int_{\tilde h_0}^{\tilde h_0(1+a)} \tilde p{\rm d}\tilde h=\tilde p_a \tilde h_0\ln(1+a)$, and the work to compress from $\tilde h_0$ to $\tilde h_0(1-a)$ is $\tilde w_{-a}=\tilde p_a\tilde h_0\ln(1-a)$, where we have used $\tilde p \tilde h=\tilde p_a \tilde h_0$ from the isothermal ideal gas law.
This gives $\lvert \tilde w_{-a}\rvert-\lvert \tilde w_{+a}\rvert=-\tilde p_a\tilde h_0\ln(1-a^2)=\tilde p_a \tilde h_0 \tilde a^2 + \OO{a^4}>0$, i.e., it is more energetically costly to expand than to compress an ideal gas by the same volume increment.
This is simply due to the fact that the isothermal bulk modulus of the gas is its pressure, which increases during compression and decreases during expansion.

We now come back to the original setting, the displacement of the rigid sheet not constrained, but controlled by a time-periodic force $\alpha \cos(t)$.
The same magnitude of force applies when compressing and expanding periodically, and we can anticipate that at each cycle, there will be more expansion than compression, i.e., the sheet will slowly move away from the wall.
An asymptotic expansion performed in \Cref{sec:asymptoticsCompressible} shows that the sheet's time averaged position follows at $\OO{\alpha^2 \Sq}$:
\begin{align}
    \frac{1}{\alpha^2\langle h \rangle ^2}\ddt{\langle h \rangle}=\frac{3}{8}\Sqbv \langle h \rangle.
\end{align}
We can compare this to \eqref{eq:gij}, where the isolated effect of gravity on a rigid sheet leads to ${\rm d}\langle h\rangle/{\rm d}t = \mathcal G \langle h \rangle ^3/4$. 
This suggests that the first-order effect of compressibility can be interpreted as increasing the effective dimensionless weight $\mathcal G_{\rm eff}$ (or dimensional weight $\tilde W_{\rm eff}$) such that:
\begin{align}
    \mathcal G_{\rm eff} = \mathcal G + \frac32 \alpha^2 \Sqbv, \quad
    \tilde W_{\rm eff} = \tilde W + \frac32 \frac{\tilde F_a^2}{\tilde p_a \tilde R^2}.
    \label{eq:effectiveWeight}
\end{align}

While we are only interested in small Squeeze numbers, we note for completeness that the averaged normal force acting on a sheet subject to a squeeze film motion can be approximated analytically for arbitrary values of ${\rm Sq}$ using the compressible (and inertialess) lubrication equations \eqref{eq:massbalance2} and \eqref{eq:fluxRe} with boundary condition $\eqref{eq:bcPressureMain}$ \citep{Taylor1957,Langlois1962}.

\subsection{Soft sheets}
\begin{figure}
	 \centering
     \includegraphics[width=\textwidth]{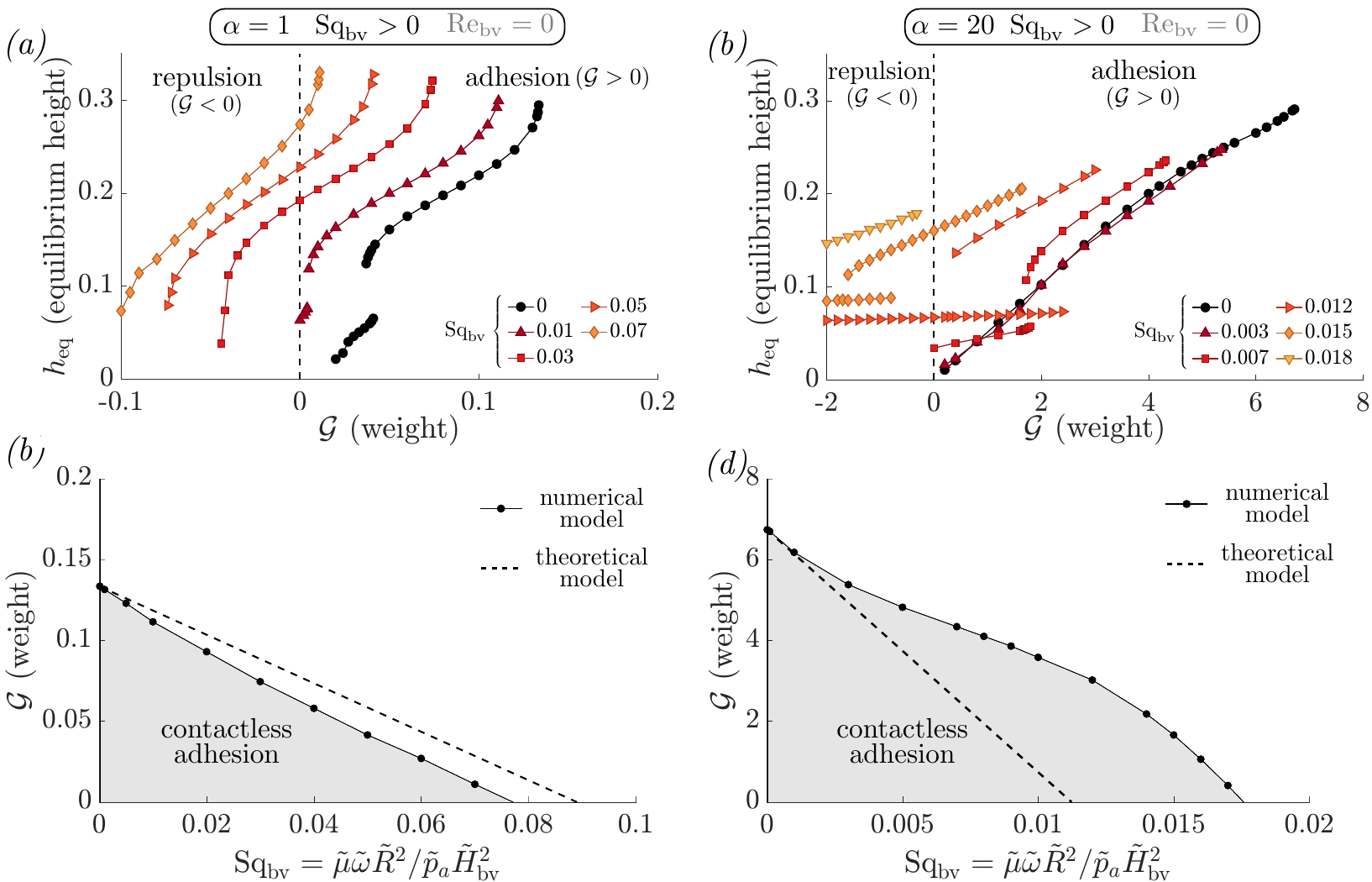}
	\caption{
    $(a-b)$ Effect of the fluid compressibility for $\Reynbv=\mathcal I_{\rm bv}=0$ and ${\rm Sq}_{\rm bv}>0$. $(a,b)$ correspond to the weak forcing regime with $\alpha=1$ and $(c,d)$ to the strong forcing regime with $\alpha=20$.
    The dashed lines in $(b)$ and $(d)$ are derived from \eqref{eq:effectiveWeight} and show $\mathcal G_{\rm max}({\rm Sq}_{\rm bv})=\mathcal G_{\rm max}({\rm Sq}_{\rm bv}=0)-1.5 \alpha^2 {\rm Sq}_{\rm bv}$.
    We primarily captured the first equilibrium branch and did not systematically investigate the entire extent of the bifurcation diagram.}
    \label{fig:SqueezeSoft}
\end{figure}

We present our results for soft sheets in the inertialess limit ($\Reynbv=\mathcal I_{\rm bv}=0$) for the weak forcing regime, $\alpha=1$, in \cref{fig:SqueezeSoft}$(a,b)$.
The results are consistent with \eqref{eq:effectiveWeight}, and the main effect of compressibility is indeed to modify the effective weight with a relation close to \eqref{eq:effectiveWeight}.
As compressibility translates the bifurcation diagrams to lower weights, we observe the possibility of reaching an equilibrium height $h_{\rm eq}>0$ for $\mathcal G<0$, corresponding to gravity pointing towards the wall. In these cases, the repulsive effect of compressibility can balance the adhesive effects of both elastohydrodynamics and gravity.
 This is reminiscent of near-field acoustic levitation \citep{Shi2019} where objects can levitate above rapidly vibrating surfaces.

We observe similar results in \cref{fig:SqueezeSoft}$(c,d)$ for the strong forcing regime with $\alpha=20$. The relation \eqref{eq:effectiveWeight} predicts $\mathcal G_{\rm max}$ accurately up to ${\rm Sq}_{\rm bv} \approx 10^{-3}$, but underestimates the effective weight by up to $50\%$ at larger squeeze numbers, likely because of the interactions between elastic deformations and compressibility neglected by this simple model. The effects on the equilibrium curves are also more complex than a simple shift of the bifurcation diagram to lower values of weight; in particular, we observe the possibility of hysteresis, similar to the weak forcing regime $\alpha \lesssim 1$ (\cref{fig:bifurcation_alpha0}$b$).

\section{Discussion and conclusion}
\label{sec:conclusion}

We have investigated numerically the elastohydrodynamic adhesion of elastic sheets vibrating near a rigid substrate.
Our analysis extends our previous work \citep{Poulain2024} by systematically exploring the regime beyond the weak-forcing limit and by incorporating the first-order corrections of fluid inertia and fluid compressibility into the viscous lubrication dynamics.
By using a combination of asymptotic analysis and numerical simulations, we have characterized the conditions under which contactless adhesion or hovering can be maintained.

The described viscous elastohydrodynamic adhesion mechanism exhibits two regimes depending on the relative strength of the forcing, $\alpha=\tilde F_a/\tilde F_{\rm bv}$ with $\tilde F_a$ the forcing amplitude and $\tilde F_{\rm bv}=(\tilde \mu \tilde \omega \tilde B^2)^{1/3}$ the elastohydrodynamic force scale involving both bending and viscous effects.
For weak forcings ($\alpha \lesssim 1$), the full dynamics can be treated analytically using asymptotic analysis, and the maximum supported weight scales as $\tilde W_{\rm max}\sim \tilde F_a^2/\tilde F_{\rm bv}$.
As the forcing strength increases, the maximum supported weight saturates and our simulations show $\tilde W_{\rm max}\sim \tilde F_{\rm bv}$ for $\alpha \gg 1$.
In both cases, the adhesion height scales as $\tilde h_{\rm eq}\sim \tilde H_{\rm bv}=\tilde R^2 (\tilde \mu \tilde \omega/\tilde B)^{1/3}$ and the sheets respond with higher-order modes of deformation as both the weight and adhesion height decrease.

In the strong forcing limit, we have identified a transition from contactless adhesion to a regime where the sheet's edges periodically come into contact with the substrate.
While the present model captures this transition numerically, a full description of the physics of contact would require additional modeling of the sheet–substrate interactions, and physical effects at a very small scale may additionally impact the dynamics. We also expect that the precise weight distribution, assumed to be uniform in our study, would start to play a significant role.

In addition to the viscous elastohydrodynamics, we have analyzed the role of fluid inertia in this lubrication flow by accounting for finite Reynolds numbers ${\rm Re}_{\rm bv}=\tilde \rho_a \tilde R^2 \tilde \omega^{2}/\tilde F_{\rm bv}$.
We have quantified how the Bernoulli-like increase of pressure due to the stagnation point in the thin gap leads to an overall destabilizing effect.
This effect decreases the maximum supported weight such that the Reynolds number based on the equilibrium height, ${\rm Re}_{\rm eq}={\tilde \rho_a \tilde \omega \tilde h_{\rm eq}^2}/{\tilde \mu}={\rm Re}_{\rm bv}(\tilde h_{\rm eq}/\tilde H_{\rm bv})^2$, remains relatively small. In particular, we predict that the sheet cannot respond solely through the first-order mode of bending deformations and that higher-order deformation modes must be excited in these experiments; a prediction that can be tested experimentally. While fluid inertia is expected to significantly influence elastohydrodynamic adhesion, our current understanding of its contribution remains qualitative. A more detailed analysis of the fluid–structure coupling at intermediate Reynolds numbers would be valuable for future work.
In particular, the range of validity of inertial corrections to lubrication theory for deformable geometries at finite Reynolds numbers, as well as the choice of appropriate boundary conditions, remains an open question.

Moreover, we considered the influence of fluid compressibility for small values of the Squeeze number ${\rm Sq}_{\rm bv}=(\tilde \mu \tilde \omega \tilde B^2)^{1/3}/\tilde p_a \tilde R^2$ and find that compressibility corrections can be interpreted as an effective weight, which modifies the adhesion threshold predicted by the incompressible analysis.
This destabilizing effect remains modest under typical experimental conditions, and our analysis suggests that compressibility is unlikely to play a significant role in the experiments of \citet{Weston2021}, in contrast to previous predictions \citep{Ramanarayanan2022,Ramanarayanan2022b}.

The present model relies on several simplifying assumptions.
In particular, it treats the deformation of the sheets as pure bending. While we do not expect qualitatively different behaviors between results in one or two dimensions for small deformations \citep{Poulain2024}, this neglects any stretching that would be important for large deformations in two dimensions and the associated possibility of non-axisymmetrical deformations when stretching becomes significant.
Second, we have neglected solid inertia. In practice however, the dimensionless parameter comparing effects of solid inertia to viscous elastohydrodynamic effects, $\mathcal I_{\rm bv}=\tilde \rho_s \tilde e \tilde \omega^2 \tilde R^4/\tilde B ={\rm Re}_{\rm bv} (\tilde \rho_s/\tilde \rho_a) \tilde e (\tilde \mu \tilde \omega/\tilde B)^{1/3}$, can be large and the role of solid inertia, along with possible resonance effects \citep{Ramanarayanan2024}, should be addressed to characterize the system fully.

In conclusion, our model provides a comprehensive framework for understanding viscous elastohydrodynamic adhesion in vibrated elastic systems, suggesting that this mechanism may operate across scales.
While our analysis has been motivated by forced centrimetric sheets, experimentally shown to support from a few hundred grams to tens of kilograms, the underlying interplay between periodic forcing, elastic deformation, and the nonlinear response of confined viscous flows is general.
We anticipate that similar principles could be leveraged in microscale systems: Applications might include tunable adhesion forces in MEMS or novel pick-and-place strategies for small objects as an alternative to standard, rigid inverted near-field acoustic levitation.
One natural avenue for future investigation is the effect of surface texture, as experiments have shown that surface roughness can be detrimental to adhesion \citep{Weston2021}. This remains to be understood, and also raises the question of whether textured surfaces can be used as a means to control adhesion and, possibly, lateral translation.
Indeed, the vibrating sheet can translate along the surface when lateral symmetry is broken \citep{Weston2021,Jia2023}, either through spatially varying forcing or gradients in material properties such as stiffness. Exploring this dynamics could allow for the design not only of contactless grippers but also of soft vibrating ``swimmers" able to hover near surfaces.
The predictions and perspectives developed herein will hopefully inspire detailed experimental investigations.

\section*{Declaration of Interests}
The authors report no conflict of interest.

\section*{Acknowledgments}
We thank Sami Al-Izzi, Annette Cazaubiel, and Jingbang Liu for insightful discussions.

\section*{Funding}
S.P. and A.C. acknowledge funding from the Research Council of Norway through project 341989. T.K. acknowledges funding from the European Union’s Horizon 2020 research and innovation programme under the
Marie Sklodowska-Curie grant agreement No 801133.
L.M. acknowledges funding from the Simons Foundation
and the Henri Seydoux Fund.

\appendix

\section{Inertial correction to lubrication theory}
\label{sec:derivationNS}
In this appendix, we consider no compressible effect and start from the incompressible Navier-Stokes equation:
\begin{align}
\tilde \rho \left(\partiald{\vect{\tilde u}}{\tilde t}+\vect{\tilde u} \cdot \vect{\tilde \nabla} \vect {\tilde u} \right)&=-\vect{\tilde \nabla} \tilde p + \tilde \mu \vect \nabla^2 \vect{\tilde u}, \\
\vect{\tilde \nabla} \cdot \vect{\tilde u} &= 0.
\label{eq:fullNS_incompressible}
\end{align}
The horizontal and vertical velocities are scaled with $\ddim \omega \ddim R$ and $\ddim \omega \ddim H$, respectively, with $\ddim H$ an arbitrary height scale.
The pressure is scaled with $\ddim \mu \ddim \omega/\varepsilon^2$ following lubrication theory, with $\varepsilon=\tilde H/\tilde R$: we expect longitudinal pressure gradients to induce a flow mainly resisted by transverse viscous stresses \cite{Batchelor}.
We make  \eqref{eq:fullNS_incompressible} dimensionless,
\begin{align}
\begin{split}
    \adim t = \ddim t\ddim\omega, \quad (\adim x, \adim y)=(\ddim x,\ddim y)/\ddim L, \quad (\adim z, \adim h)=(\ddim z,\ddim h)/\ddim H, \\
    \quad \adim{\vect v}_\perp=\ddim{\vect v}_\perp/\ddim \omega \ddim R,  \quad \adim v_z =\ddim v_z/\ddim \omega \ddim H, \quad     \adim p = \varepsilon^2 \ddim p / (\ddim \mu \ddim\omega), \quad \rho = \tilde \rho/\tilde \rho_a,
\end{split}
\end{align}
and assume $\varepsilon=\tilde H/\tilde R\ll 1$, ${\Reyn}=\tilde \rho_a \tilde H^2 \tilde \omega/\tilde \mu\ll 1$. We also differentiate between the horizontal direction $\xperp=(x,y)$ and the vertical direction $z$:
\begin{subequations}
\begin{align}
    \Reyn \left( \partiald{}{\adim t}+\left(\vect{\adim v} \cdot \adim\grad\right)\right)\vect{\adim v}_\perp &= -\adim{\vect \grad}_\perp p +  \partialdd{\vect{\adim v}_\perp}{\adim z}
    + \OO{\varepsilon^2}, 
    \label{eq:NS_scaled}\\
    0 &= -\partiald{p}{\adim z}     +  \OO{\varepsilon^2,\varepsilon^2 \Reyn}
    \label{eq:buoy_zero}, \\
     \adim \grad_\perp \cdot \adim{\vect v}_\perp + \partiald{\adim v_z}{\adim z}  &=0.
    \label{eq:continuity}
\end{align}
    \label{eq:fullscaledNS}
\end{subequations}

We seek a depth-integrated description of the flow. The momentum balance normal to the wall \eqref{eq:buoy_zero} shows that $\tilde p$ is independent of $\ddim z$. Integrating \eqref{eq:continuity} from the wall at $\tilde z=0$ to the sheet at $\tilde z=\tilde h$, and applying Leibniz integral rule together with the kinematic boundary condition $\partial  h/\partial t + \vperp \gradperp h = v_z\lvert_{z=h} $ yields
\begin{equation}
\begin{split}
	\partiald{h}{t}	+ \gradperp \cdot \vect q=0,
\label{eq:governing_mass}
\end{split}
\end{equation}
with $ \vect q= \int_0^h \vperp~{\rm d}z =  \ddim{\vect q}/\ddim \omega \ddim L \ddim h $ the volumetric flux along the wall.

We then follow ideas from \cite{Rojas2010}, who included the effects of inertia at first order in the Reynolds number to study thin liquid films: we adapt their derivation to consider a solid surface instead. 
 We start with a Taylor expansion of the velocity and pressure in the normal direction:
\begin{equation}
\begin{split}
    \vperp (\vect x,t)&=\sum_{n=0}^{+\infty} \vect v_n(\xperp,t)\frac{z^{n+1}}{(n+1)!}, \\
    v_z (\vect x,t)&=-\sum_{n=0}^{+\infty}\left(\gradperp\cdot \vect v_n\right)(\xperp,t)\frac{z^{n+2}}{(n+2)!}, \\
   p (\vect x,t)&=\sum_{n=0}^{+\infty} p_n(\xperp,t)\frac{z^{n}}{n!}.
\end{split}
    \label{eq:TaylorNS}
\end{equation}
We can write the first expression thanks to the no-slip condition $\vperp=0$ at $z=0$, while the second expression comes from both the no-penetration condition $v_z=0$ at $z=0$ and the continuity equation \eqref{eq:continuity}.
Inserting \eqref{eq:TaylorNS} into \eqref{eq:fullscaledNS} and identifying the coefficients of the Taylor series yields:
\begin{equation}
\begin{split}
    p_{n\geq 1} &= \mathcal{O}(\varepsilon^2,\varepsilon^2\Reyn), \\
    \vect v_1 &= \gradperp p_0, \\
    \vect v_2 &= \Reyn \partiald{\vect v_0}{t} + \mathcal{O}(\varepsilon^2),   \\
    \vect v_3 &= \Reyn \left[\partiald{\left(\gradperp p_0\right)}{t} + 2\left(\vect v_0 \cdot \gradperp\right)\vect v_0 - \vect v_0\left(\gradperp\cdot \vect v_0\right)\right] +\mathcal{O}(\varepsilon^2),\\
    \vect v_4 &= \Reyn \left[3\left(\gradperp p_0\cdot\gradperp\right)\vect v_0 - 3\left(\gradperp \cdot \vect v_0\right)\gradperp p_0 + 3 \left(\vect v_0\cdot \gradperp\right)\gradperp p_0 - \right. \\& \qquad \qquad  \left. - \vect v_0\left(\gradperp^2 p_0\right) \right] +\mathcal{O}\left(\varepsilon^2,\varepsilon^2\Reyn,\Reyn^2\right), \\
    \vect v_5 &= \Reyn\left[-4\gradperp p_0 \left(\gradperp^2p_0\right) + 6\left(\gradperp p_0 \cdot \gradperp\right)\gradperp p_0\right] + \mathcal O\left(\varepsilon^2,\varepsilon^2\Reyn,\Reyn^2\right), \\
    \vect v_{n \geq 6} &= \mathcal O\left(\varepsilon^2,\varepsilon^2\Reyn,\Reyn^2\right).
\end{split}
    \label{eq:TaylorSpeed}
\end{equation}
We seek $\vect v_0$ and $p_0$ and expand them in powers of the Reynolds number:
\begin{equation}
\begin{split}
    \vect v_0 &= \vect v_0^{(0)} + \Reyn~\vect v_0^{(1)} + \mathcal O(\Reyn^2), \\
    p_0 &=  p_0^{(0)} + \Reyn~p_0^{(1)} + \mathcal O(\Reyn^2).
\end{split}
\end{equation}

Using the definition of the volumetric flux $\vect q$ and evaluating the no-slip condition $\vperp=0$ at $z=h$, we find:
\begin{equation}
\begin{split}
    \vect q =& \left[\vect v_0^{(0)} \frac{h^2}{2!} + \gradperp p_0^{(0)} \frac{h^3}{3!}\right]
    +  \\& \Reyn \left[
    \vect v_0^{(1)} \frac{h^2}{2!} + \gradperp p_0^{(1)} \frac{h^3}{3!} + \vect v_2^{(1)} \frac{h^4}{4!} + \vect v_3^{(1)} \frac{h^5}{5!} + \vect v_4^{(1)} \frac{h^6}{6!} + \vect v_5 ^{(1)}\frac{h^7}{7!}
    \right], \\
    0 =& \left[\vect v_0^{(0)}h + \gradperp p_0^{(0)} \frac{h^2}{2!}\right]
    + \\& \Reyn \left[
    \vect v_0^{(1)} h + \gradperp p_0^{(1)} \frac{h^2}{2!} + \vect v_2^{(1)} \frac{h^3}{3!} + \vect v_3^{(1)} \frac{h^4}{4!} + \vect v_4^{(1)} \frac{h^5}{5!} + \vect v_5^{(1)} \frac{h^6}{6!}
    \right].
\end{split}
\end{equation}

We can solve this linear system order by order to express $\vect v_0$ and $\gradperp p_0$ as a function of $\vect q$ and $h$:
\begin{align}
\begin{aligned}
    \vect v_0^{(0)} &= \frac{6 \vect q}{h^2}, \\
    \gradperp p_0^{(0)} &= -\frac{12 \vect q}{h^3},
    \label{eq:pressflux} \\
    \vect v_0^{(1)} &=  \frac{h^2}{12}\vect v_2^{(1)} + \frac{h^3}{30}\vect v_3^{(1)} + \frac{h^4}{120}\vect v_4^{(1)} + \frac{h^5}{630}\vect v_5^{(1)}
    , \\
    \gradperp p_0^{(1)} &=  -\frac{h}{2}\vect v_2^{(1)} - \frac{3h^2}{20}\vect v_3^{(1)} - \frac{h^3}{30}\vect v_4^{(1)} - \frac{h^4}{168}\vect v_5^{(1)}
    ,
\end{aligned}
\end{align}
where the $v_n$, $n\geq 2$, are found from \eqref{eq:TaylorSpeed}.
After lengthy calculations, we find the contribution at $\mathcal O (\Reyn)$ of the pressure gradient as:
\begin{align}  
\gradperp p_0^{(1)}=
     -\frac65 \partiald{}{t}\left(\frac{\vect q}{h}\right)
     - \frac{54}{35} \frac{\vect q}{h} \cdot \gradperp\left(\frac{\vect q}{h}\right)
     + \frac6{35} \frac{\vect q}{h^2}\partiald ht,
\end{align}
which is the same as for the case of a free surface \citep{Rojas2010}, even though the velocity profile \eqref{eq:TaylorSpeed} differs.
\Cref{eq:pressflux} finally gives the link between the velocity flux $\vect q$ flux and the pressure gradient as:
\begin{subequations}
\begin{align}
    12 \vect q +  h^3\gradperp p + \frac65\Reyn h^3 \left(\partiald{}{t}\left(\frac{\vect q}{h}\right)
     + \frac{9}{7} \frac{\vect q}{h} \cdot \gradperp\left(\frac{\vect q}{h}\right)
     - \frac1{7} \frac{\vect q}{h^2}\partiald ht\right)  = 0,
    \label{eq:inertiaThin}
\end{align}
or, in dimensional units:
\begin{align}
    12\ddim \mu \ddim{\vect q} +  \ddim h^3 \ddim \grad_\perp\ddim p + \frac65\ddim \rho \ddim h^3 \left(\partiald{}{\ddim t}\left(\frac{\ddim{\vect q}}{\ddim h}\right)
     + \frac{9}{7} \frac{\ddim{\vect q}}{\ddim{h}} \cdot \gradperp\left(\frac{\ddim{\vect q}}{\ddim{h}}\right)
     - \frac1{7} \frac{\ddim{\vect q}}{\ddim{h}^2}\partiald{\ddim h}{\ddim t}\right)  = 0.
    \label{eq:inertiaThinDim}
\end{align}
\end{subequations}
This recovers the analyses of \citet{Kuzma1968,Tichy1970,Jones1975} when $h$ is the distance between two rigid, horizontal plates with then $h=h(t)$, $q(x,t)=-(x/h){\rm d }h/{\rm d}t$ in 1D and $q(r,t)=-(r^2/2h){\rm d }h/{\rm d}t$ in 2D axisymmetric with $r$ the radial coordinate.
This also recovers the first-order correction of \cite{Ishizawa1966} when, in addition, $h(t)$ oscillates sinusoidally. 

\section{Pressure boundary condition}
\label{sec:Pressurebc}
Without fluid inertia, for $\Reynbv=0$, we can impose the pressure at the edge of the sheet to match the ambient pressure: $p=0$.
There is, in fact, a non-slender region near the edges connecting the gap to the outside where lubrication theory might break down.
There, we assume that the pressure difference scales as $\Delta \ddim p\sim \ddim \mu  \ddim v_\perp/\ddim L \sim \ddim \mu \ddim \omega$ with $\tilde v_\perp=\tilde \omega \tilde L$ the horizontal velocity scale.
Comparing this to the dynamic pressure that scales as, we find $\ddim \rho_a (\ddim \omega \hbv)^2/\Delta\ddim p \sim \Reynbv$, which suggests a boundary effect at $\OO{\Reynbv}$.

This problem has been discussed in prior works.
For the steady translation of a rigid sheet, \citet{Tuck1983} argue that when fluid leaves the gap the pressure at the edge matches the ambient pressure. This is a classical condition for a jet.
However, when fluid enters the gap, the entrance flow is not jet-like, and fluid is drawn from an extended region, leading to a pressure drop. 
Significant effects of inertia on the pressure distribution can arise due to this sole asymmetry in entrance and exit boundary conditions \citep{Tuck1983,Tichy1985}.
For thin film flows between an oscillating plate and a stationnary wall, the early study of \citet{Kuroda1976} has been extensively followed \citep{Hori2006} and matches the work of \citet{Tuck1983}: at the edges, the author assumes $p=0$ when $\vect q\cdot \vect{e_r}>0$ (outflow) and $\tilde p=-(k/2)\ddim \rho_a ( \ddim{\vect q} \cdot \vect{e_r} / \ddim h )^2$ otherwise (inflow). Here, $\ddim{\vect q} \cdot \vect{e_r}/\ddim h$ is the average velocity of the fluid leaving or entering the gap, and $k>0$ is a dimensionless coefficient taking into account losses in the Bernoulli pressure drop.
In dimensionless quantities, this boundary condition reads:
\begin{align}
    p=
    \begin{cases}
     0~&\text{if}~\vect q \cdot \vect e_r > 0~~\text{(outflow)} \\
     - \dfrac{k}{2} \Reynbv \left(\dfrac{\vect q \cdot \vect e_r }{h}\right)^2~&\text{if}~\vect q \cdot \vect e_r < 0 ~~\text{(inflow)}
    \end{cases}.
    \label{eq:bcPressure}
\end{align}
The value $k=0.5$ is often adopted \citep{Kuroda1976,Hori2006} by analogy with high-Reynolds-number pipe flows \citep{Cengel2013}.

Recently, \cite{Ramanarayanan2022} studied in detail the effect of inertia for the flow under a flat and circular rigid plate undergoing oscillations above a solid substrate as 
$\tilde h(\tilde t)=\tilde h_0 ( 1+a\sin(\tilde \omega \tilde t) )$, $0<a<1$.
They matched the thin film flow in the gap to solutions of the Navier-Stokes equations outside the gap for a wide range of Reynolds numbers $\Reyn=\tilde \rho_a \tilde \omega \tilde h_0/\tilde \mu$.
They found that the pressure at the edges averaged over one period of oscillation is $\langle \ddim p_e \rangle=-K \ddim \rho_a \ddim R^2 a^2 \ddim \omega^2$, with $K$ a coefficient found numerically.
For $\Reyn\lesssim 5$, $K \approx 0.096$. For $\Reyn\gtrsim 100$, $K=1/16$ if  $(\ddim H/\ddim R) \ll a$, $K=1/32$  if  $(\ddim H/\ddim R) \gg a$. 
The boundary condition \eqref{eq:bcPressure} applied to the same situation yields the same scaling: $\langle \ddim p_e \rangle=-(k \pi/16) \ddim \rho \ddim L^2 a^2 \ddim \omega^2$. This matches the various cases studied by \cite{Ramanarayanan2022} if $k\simeq0.49$, $k\simeq 0.32$ and $k \simeq 0.16$, respectively.
Since we will be dealing with Reynolds numbers that remain small, $\mathrm{R_e}< 10 $, we adopt \eqref{eq:bcPressure} with $k=0.5$.

\section{Viscous adhesion under weak active forcing}
\label{sec:appendixWeak}

To study theoretically \eqref{eq:Reynolds} with the force balance \eqref{eq:normalForceBalance_dimensionless}, we use a Galerkin projection of the height. We let
\begin{subequations}
    \begin{align}
    h(x,t)=h_0(t)+\mathcal G H_1(x) + \alpha \cos(t) H_0(x) + \alpha \sum_{i=1}^{\infty}a_i(t)\zeta_i(x),
\end{align}
where $h_0(t)$ and $(a_i(t))_{i \in \mathbb N^\star}$ are unknown time-dependent coefficients.
The functions $H_0$, $H_1$ and $(\zeta_i)_{i \in \mathbb N^\star}$ are chosen as:
\begin{align}
    H_{0}(x)&=-\frac{x^6}{240}+\frac{x^4}{16}-\frac{\lvert x\rvert^3}{6}+\frac{3x^2}{16}, \quad  H_{1}(x)=-\frac{x^6}{240}+\frac{x^4}{48}-\frac{x^2}{16}, \label{eq:ansatz}\\
    \zeta_n(x)&=I_n \times
    \begin{cases}
         (-1)^{\frac n2}\cosh\left(\frac{\sqrt 3}{2}n\pi\right)\cos(n\pi x) + 2\cosh\left(n\pi \frac{\sqrt{3}}{2}x\right)\cos\left(\frac{n\pi}2x\right), ~n~\mathrm{even}\\
       (-1)^{\frac{n-1}{2}}\sinh\left(\frac{\sqrt 3}{2}n\pi\right)\cos(n\pi x)-2\sinh\left(n\pi \frac{\sqrt{3}}{2}x\right)\sin\left(\frac{n\pi}2x\right), ~n~\mathrm{odd},
    \end{cases} \nonumber
\end{align}
\label{eq:ansatz}
\end{subequations}
with $I_n$ a normalization coefficient ensuring that $\int_0^1 \zeta_n^2=1$.
 This expansion recovers the first-order deformation $H_1$ and $H_0$ due to a uniform load and to a point-active forcing, respectively, while allowing for higher-order modes of deformation.
 The $\zeta_n$, shown in \cref{fig:bifurcation_alpha0}$(d)$, are the even eigenmodes of the triharmonic operator $\partial/\partial x^6$ satisfying the boundary conditions $\partial^2 \zeta_n/\partial x^2=\partial^3 \zeta_n/\partial x^3=\partial^4 \zeta_n/\partial x^4=0$ at $x=\pm 1$, such that this ansatz for $h$ satisfies the boundary conditions \eqref{eq:bc_elastic} and \eqref{eq:bcPressureMain}.

 Upon inserting \eqref{eq:ansatz} in the governing equations and projecting in space, we make use of an averaging method considering a separation of timescales between the fast oscillation time $t$ and the slow time $\alpha^2 t$ associated with the long-time evolution of the system (\cref{fig:schematic}$b$).
This gives rise to an evolution equation at $\OO{\alpha^2}$ for $\langle h_0 \rangle(t) = \int_t^{t+2\pi}h(x=0,t')~{\rm d}t'$, the time-averaged evolution of the sheet's center height, given by \eqref{eq:gij}.
 
Up to $N=5$, the coefficients $d_{ij}$ appearing in \eqref{eq:gij} are found numerically and are given by the entries of the following symmetric matrix
\begin{align}
\mat d=\begin{pmatrix}
\num{1.14e-2} & \num{2.04e-6} & \num{-6.91e-6}  & \num{1.64e-5} & \num{-3.24e-5 }\\
 & \num{6.95e-4} & \num{9.24e-7} & \num{-2.05e-6} & \num{4.01e-6}  \\
 &  & \num{1.49e-4} & \num{7.08e-7} & \num{-1.24e-6}  \\
  &  & & \num{3.00e-5} & \num{6.12e-7}  \\
    &  & & & \num{4.73e-5}  \\
\end{pmatrix},
\end{align}
\label{eq:asymptoticsappendix}
and $d_0=0.122$.

\begin{figure}
	 \centering
\includegraphics[width=\textwidth]{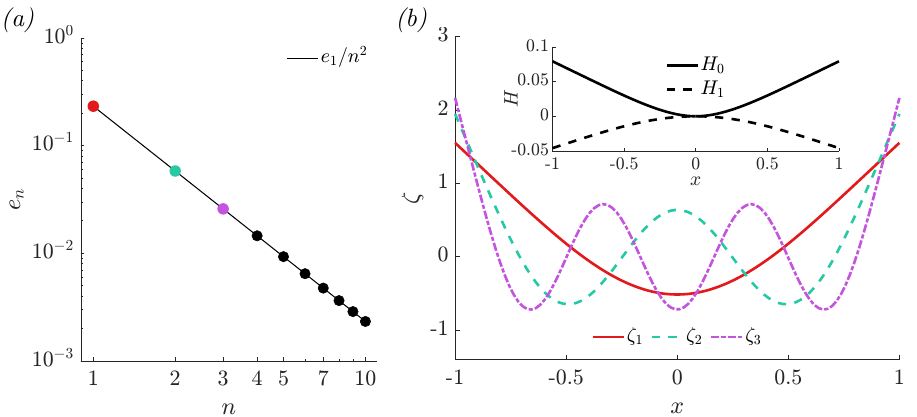}
	\caption{\label{fig:bifurcation_alpha0_new22}
Asymptotic results for $\alpha \lesssim 1$, $\mathcal I_{\rm bv}=\Reynbv=\Sqbv=0$, adapted from \citet{Poulain2024}.
As $\mathcal G$ and $h_{\rm eq}$ decrease, the sheet presents higher and higher order deformation modes. The $i-$th mode $\zeta_i$ is excited if $h \lesssim e_i$.
}
\end{figure}

\section{Rigid sheet and non-zero Reynolds number}
\label{sec:inertia_asymp}
We consider the case of a rigid and weightless  1-dimensional sheet in an incompressible fluid in order to isolate the effects of fluid inertia: $\tilde B \rightarrow \infty,~\mathcal G=0,~\Sq=0,~\Reyn>0$.
The governing equations reduce to:
\begin{subequations}
\begin{align}
	&12\partiald ht - \partiald{}{x}\left(h^3\partiald px\right)=0, \\
	&\int_{-1}^{+1}p~{\rm dx}=-2\alpha \cos(t),
	\label{eq:reyn222}
\end{align}
\label{eq:reyn111}
with the boundary condition at $x=\pm 1$:
	\begin{align}
\begin{aligned}
	 p&=
    \begin{cases}
     0~&\text{if}~q>0 \\
     - \dfrac{k}{2} \Reyn \left(\dfrac qh\right)^2~&\text{if}~q< 0
    \end{cases}.
\end{aligned}
\end{align}
\end{subequations}

We consider a sheet aligned with the wall, $h=h(t)$.
We can express the pressure gradient as a function of the height $h$ from \eqref{eq:reyn111}:
\begin{align}
    \frac 1x \partiald px = 12 \frac{\dot h}{h^3} + \frac{6\Reyn}{5} \frac{\ddot h}{h} - \frac{102\Reyn}{35}\frac{\dot h^2}{h^2}.   
\end{align}
This can directly be integrated using the boundary condition to give the pressure profile in the gap:
\begin{align}
    p(x,t)&=\left(1-x^2\right)\left(-6 \frac{\dot h}{h^3} - \frac{3\Reyn}{5} \frac{\ddot h}{h} + \frac{51\Reyn}{35}\frac{\dot h^2}{h^2}\right) + p_{\rm edge}, \\ 
    p_{\rm edge}&=-\frac k2 \Reyn \heaviside{\dot h} \left(\frac{\dot h}{h}\right)^2,
    \label{eq:press_reyn}
\end{align}
with $\mathbb H$ the Heaviside function.
Using \eqref{eq:reyn222}, this leads to the following ordinary differential equation for $h(t)$, together with the initial conditions $h(0)=1,~\dot h(0)=0$:
\begin{align}
 \frac{2\Reyn}{5h} \ddot h + 4 \frac{\dot h}{h^3} + \Reyn \frac{\dot h^2}{h^2} \left[ \frac k2\heaviside{\dot h} - \frac{34}{35} \right] -\alpha \cos(t)=0.
  \label{eq:reynoldsODE}
 \end{align}
We note that the unsteady inertial term, proportional to $\ddot{h}$, appears as an added mass as mentioned in \S \ref{sec:inertiaRigidSheet}.

To find an approximate solution to   \eqref{eq:reynoldsODE}, we introduce a slow timescale $\tau_{\rm Re}=\Reyn t$ and expand the height in powers of the Reynolds number:
\begin{align}
		h(t)=h_0(t,\tau_{\rm Re})+\Reyn h_1(t,\tau_{\rm Re})+\OO{\Reyn^2}.
		\label{eq:asymp_re}
\end{align}
Inserting into \eqref{eq:reynoldsODE} yields:
\begin{subequations}
\begin{align}
\OO1:& \quad
	4\partiald{h_0}{t}+\alpha \cos(t)h_0^3=0,
	\label{eq:Re1} \\
	\OO{\Reyn}:& \quad
	4\partiald{h_1}{t}+3h_0^2\alpha \cos(t)h_1= \nonumber\\ &\qquad\qquad -4\partiald{h_0}{\tau_{\rm Re}}-\frac25h_0^2\partialdd{h_0}{t}+h_0\left(\partiald{h_0}{t}\right)^2
	\left[\frac{34}{35}-\frac k2 \heaviside{\dot h_0} \right].
	\label{eq:Re2} 
\end{align}
\end{subequations}
Equation \eqref{eq:Re1} can be integrated directly as a linear differential equation in $t$. Introducing $f_{\Reyn}({\tau_\Reyn})$ as an integration constant (a function of $\tau_\Reyn$  independent of $t$), we find:
\begin{align}
	h_0=\left[f_{\rm Re}(\tau_{\rm Re})+\frac{\alpha}2\sin(t)\right]^{-1/2}.
\end{align}
Knowing $h_0$, \eqref{eq:Re2} is also a linear ordinary differential equation for $h_1$ which can be solved as:
\begin{align}
\begin{split}
    h_1(t,\tau_{\rm Re}) = h_0^3(t) \int_0^t \bigg[ 
        -\frac{1}{h_0^3} \partiald{h_0}{\tau_{\rm Re}}
        - &\frac{1}{10 h_0} \partialdd{h_0}{t} \\
        &+ \frac{1}{h_0^2} \left( \partiald{h_0}{t} \right)^2 
        \left( \frac{17}{70} - \frac{k}{8} \mathbb{H}\left( \partiald{h_0}{t} \right) \right)
    \bigg]\,{\rm d}t'.
    \label{eq:h1_re}
\end{split}
\end{align}
Both $h_0$ and the integrand of \eqref{eq:h1_re} are $2\pi-$periodic in $t$. Therefore, if the average value of the integrand was non-zero, $h_1(t)$ would diverge as $t\rightarrow \infty$.
This would break the asymptotic expansion \eqref{eq:asymp_re}.
We conclude that this integrand must have a zero mean, which is equivalent to requesting that $h_1$ must be $2\pi$-periodic in $t$.
In particular, $h_1(0,\tau_{\rm Re})=h_1(2\pi,\tau_{\rm Re})$. This non-secularity condition  gives a differential equation for $f_{\rm Re}(\tau_{\rm Re})$:
\begin{equation}
\begin{split}
	\pi f'_{\rm Re}(\tau_{\rm Re})=\frac{\alpha}{40} &\int_0^{2\pi}\frac{\sin(t)}{f_{\rm Re}(\tau_{\rm Re})+\frac{\alpha}{2}\cos(t)}{\rm d}t 
	\\ & + \frac{\alpha^2}{280}\int_0^{2\pi}\frac{\cos^2(t)}{\left(f_{\rm Re}(\tau_{\rm Re})+\frac{\alpha}{2}\sin(t)\right)^2}{\rm d}t
	\\  &  + \frac{k\alpha^2}{128}\int_{\pi/2}^{3\pi/2}\frac{\cos^2(t)}{\left(f_{\rm Re}(\tau_{\rm Re})+\frac{\alpha}{2}\sin(t)\right)^2}{\rm d}t.
\end{split}
\end{equation}
We approximate the integrals using a first-order Taylor expansion for small $\Gamma/2f_{\rm Re}$ and find:
\begin{equation}
\begin{aligned}
	f^2_{\rm Re}(\tau_\Reyn)f'_{\rm Re}(\tau_\Reyn)&=-\frac{1}{112}\left(1-\frac{7k}{16}\right)\alpha^2,\\
	 f_{\rm Re}(\tau_{\rm Re})&=\left(1-\frac{3}{112}\left(1-\frac{7k}{16}\right)\alpha^2\tau_{\rm Re}\right)^{1/3}, \\
     h_0(t)&=\left[\left(1-\frac{3}{112}\left(1-\frac{7k}{16}\right)\alpha^2\Reyn t\right)^{1/3} + \frac{\alpha}{2}\sin(t)\right]^{-1/2}.
\end{aligned}
\end{equation}

\begin{figure}
	 \centering
     \includegraphics[width=\textwidth]{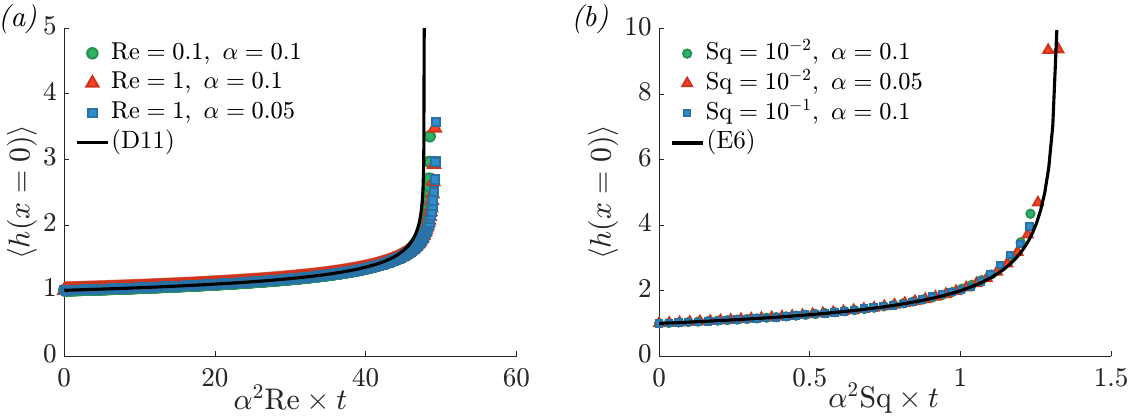}
	\caption{\label{fig:label}
    Comparison between numerical results (symbols) and first-order asymptotic calculations from two-timescale analysis (lines) for the height evolution of a rigid sheet with
    $(a)$ ${\rm Re}>0$, ${\rm Sq=0}$ and $(b)$ ${\rm Sq}>0$, ${\rm Re=0}$.
	}
\end{figure}

We have kept an arbitrary value for the loss coefficient $k$ for completeness. We argue in \Cref{sec:Pressurebc} for $k=0.5$, and with this value  $3(1-7k/16)/112 \simeq 0.021$.
Therefore, $\langle h\rangle (t) \simeq \left(1-0.021 \alpha^2\Reyn t\right)^{-1/6}$, where we denote $\langle h\rangle (t)= \int_{t'}^{t'+2\pi}h(t'){\rm d}t'/2\pi$ the time-averaged evolution.
These results are verified in \cref{fig:label}$(a)$.

\section{Rigid sheet and non-zero Squeeze number}
\label{sec:asymptoticsCompressible}
We neglect inertia, $\mathrm{R_e}=0$, but assume a non-zero squeeze number $\Sq>0$. The mass conservation equation \eqref{eq:massbalance2} and the momentum balance \eqref{eq:fluxRe} give:
\begin{align}
	12\partiald{\left(\rho h\right)}{t}-\frac{\partial}{\partial x}\left(\rho h^3 \partiald p x \right), \quad \rho=1+\Sq p.
	\label{eq:compressiblethinFilm}
\end{align}
We let $\tau_{\rm Sq}=\Sq t$ and expand  $h(t)=h_0(t,\tau_{\rm Sq}) + \Sq h_1(t,\tau_{\rm Sq})+ \OO{\Sq^2}$ and $p(x,t)=p_0(x,t,\tau_{\rm Sq}) + \Sq p_1(x,t,\tau_{\rm Sq}) + \OO{\Sq^2}$. Collecting terms of the same order, \eqref{eq:compressiblethinFilm} yields
\begin{subequations}
\begin{align}
	 \OO{1}: 	\quad &12\partiald{h_0}{t}-h_0^3\partialdd{p_0}{x}=0, \\
	 \OO{\Sq}:	\quad &12\partiald{h_1}{t}-h_0^3\partialdd{p_1}{x}=  \nonumber\\
     &\qquad -12\partiald{\left(p_0h_0\right)}{t}-12\partiald{h_0}{\tau_{\rm Sq}}+3h_0^2h_1\partialdd{p_0}{x}+h_0^3\frac{\partial}{\partial x}\left(p_0\partiald{p_0}{x}\right),
     \label{eq:squeezesth}
\end{align}
\end{subequations}
while the integrated version of the force balance \eqref{eq:normalForceBalance_dimensionless} and the boundary condition \eqref{eq:bcPressureMain} yield
\begin{footnotesize}
\begin{align}
	 \int_{-1}^1 p_0~{\rm d}x=2\alpha\cos(t), \quad \int_{-1}^{+1} p_1~{\rm d}x=0, \quad p_0(x=\pm 1,t)=p_1(x=\pm 1,t)=0.
\end{align}
\end{footnotesize}

The problem at $\OO{1}$ is simply the rigid incompressible problem, and the solution is directly found as
\begin{align}
	p_0=\frac{\alpha}{4}\cos(t)(1-x^2), \quad h_0=\left(f_{\rm Sq}(\tau_{\rm Sq})+\frac{\alpha}{2}\sin(t)\right)^{-1/2},
\end{align}
with $f_{\rm Sq}(\tau_{\rm Sq})$ an integration constant that depends on the slow time.
At $\OO{\Sq}$ we can integrate \eqref{eq:squeezesth} to obtain $p_1$, then use the force balance to obtain a single ordinary differential equation for $h_1$:
\begin{footnotesize}
\begin{align}
	\partiald{h_1}{t}+\frac{3\alpha\cos(t)}{4\left(1+\alpha\sin(t)/2\right)}h_1=\frac{20f_{\rm Sq}'(\tau_{\rm Sq})+30\alpha^2-18\alpha^2\cos(2t)+96\alpha f_{\rm Sq}(\tau_{\rm Sq})\sin(t)}{80\left(1+\alpha\sin(t)/2\right)^{3/2}} .
\end{align}
\end{footnotesize}
Using a non-secularity condition similar to the one used in \cref{sec:inertia_asymp}, we find:
\begin{align}
	f_{\rm Sq}'(\tau_{\rm Sq})=-\frac{3\alpha^2}{4},\quad f_{\rm Sq}(\tau_{\rm Sq})=1-\frac{3\alpha^2}{4}\tau_{\rm Sq}, \quad h_0(t)\approx\left(1-\frac34 \alpha^2 \Sq t \right)^{-1/2},
\end{align}
which is verified in \cref{fig:label}$(b)$.
The pressure at $\OO{{\rm S_q}}$ is
\begin{align}
	p_1(x,t,\tau)=\frac{3\alpha}{20}\left(1-x^2\right)\left(5x^2-1\right)\left[\alpha\cos(2t)-2f_{\rm Sq}(\tau_{\rm Sq})\sin(t)\right].
 \end{align}

\bibliographystyle{jfm}
\bibliography{./biblio}

\begin{thebibliography}{77}
\expandafter\ifx\csname natexlab\endcsname\relax\def\natexlab#1{#1}\fi
\def\au#1{#1} \def\ed#1{#1} \def\yr#1{#1}\def\at#1{#1}\def\jt#1{\textit{#1}} \def\bt#1{#1}\def\bvol#1{\textbf{#1}} \def\vol#1{#1} \def\pg#1{#1} \def\publ#1{#1}\def\arxiv#1{#1}\def\org#1{#1}\def\st#1{\textit{#1}}

\bibitem[Alexander(1977)]{Alexander1977}
{\sc \au{Alexander, R.}} \yr{1977}  \at{Diagonally implicit {R}unge-{K}utta methods for stiff {O.D.E.}'s}.  \jt{SIAM Journal on Numerical Analysis}  \bvol{14}~(6),  \pg{1006--1021}.

\bibitem[Andrade {\em et~al.\/}(2018)Andrade, P{\'e}rez \& Adamowski]{Andrade2018}
{\sc \au{Andrade, M.A.B.}, \au{P{\'e}rez, N.} \& \au{Adamowski, J.C.}} \yr{2018}  \at{Review of progress in acoustic levitation}.  \jt{Braz. J. Phys.}  \bvol{48},  \pg{190--213}.

\bibitem[Andrade {\em et~al.\/}(2020)Andrade, Ramos, Adamowski \& Marzo]{Andrade2020}
{\sc \au{Andrade, M.A.B.}, \au{Ramos, T.S.}, \au{Adamowski, J.C.} \& \au{Marzo, A.}} \yr{2020}  \at{Contactless pick-and-place of millimetric objects using inverted near-field acoustic levitation}.  \jt{Appl. Phys. Lett.}  \bvol{116}~(5).

\bibitem[Argentina {\em et~al.\/}(2007)Argentina, Skotheim \& Mahadevan]{Argentina2007}
{\sc \au{Argentina, M.}, \au{Skotheim, J.} \& \au{Mahadevan, L.}} \yr{2007}  \at{Settling and swimming of flexible fluid-lubricated foils}.  \jt{Phys. Rev. Lett.}  \bvol{99}~(22),  \pg{224503}.

\bibitem[Atalla {\em et~al.\/}(2023)Atalla, Van~Ostayen, Sakes \& Wiertlewski]{Atalla2023}
{\sc \au{Atalla, M.A.}, \au{Van~Ostayen, R.A.J.}, \au{Sakes, A.} \& \au{Wiertlewski, M.}} \yr{2023}  \at{Incompressible squeeze-film levitation}.  \jt{Appl. Phys. Lett.}  \bvol{122}~(24).

\bibitem[Bao \& Yang(2007)]{Bao2007}
{\sc \au{Bao, M.} \& \au{Yang, H.}} \yr{2007}  \at{Squeeze film air damping in {MEMS}}.  \jt{Sens. Actuator A-Phys.}  \bvol{136}~(1),  \pg{3--27}.

\bibitem[Batchelor(1967)]{Batchelor}
{\sc \au{Batchelor, G.K.}} \yr{1967} {\em An Introduction to Fluid Dynamics\/}.  \publ{Cambridge University Press}.

\bibitem[Bigan {\em et~al.\/}(2024)Bigan, Liz{\'e}e, Pascual, Nigu{\`e}s, Bocquet \& Siria]{Bigan2024}
{\sc \au{Bigan, N.}, \au{Liz{\'e}e, M.}, \au{Pascual, M.}, \au{Nigu{\`e}s, A.}, \au{Bocquet, L.} \& \au{Siria, A.}} \yr{2024}  \at{Long range signature of liquid's inertia in nanoscale drainage flows}.  \jt{Soft Matter}  \bvol{20}~(44),  \pg{8804--8811}.

\bibitem[Brennen(1982)]{Brennen1982}
{\sc \au{Brennen, C.E.}} \yr{1982}  \at{A review of added mass and fluid inertial forces}.  \jt{Tech. Rep. CR 82.010} Naval Civil Engineering Laboratory.

\bibitem[Bureau {\em et~al.\/}(2023)Bureau, Coupier \& Salez]{Bureau2023}
{\sc \au{Bureau, L.}, \au{Coupier, G.} \& \au{Salez, T.}} \yr{2023}  \at{Lift at low {R}eynolds number}.  \jt{Eur. Phys. J. E}  \bvol{46}~(11),  \pg{111}.

\bibitem[Calisti {\em et~al.\/}(2017)Calisti, Picardi \& Laschi]{Calisti2017}
{\sc \au{Calisti, M.}, \au{Picardi, G.} \& \au{Laschi, C.}} \yr{2017}  \at{Fundamentals of soft robot locomotion}.  \jt{J. R. Soc. Interface.}  \bvol{14}~(130),  \pg{20170101}.

\bibitem[Colasante(2015)]{Colasante2015}
{\sc \au{Colasante, D.A.}} \yr{2015} Apparatus and method for orthosonic lift by deflection. {U.S.} {P}atent {US8967965B1}.

\bibitem[Colasante(2016)]{ColasanteYT}
{\sc \au{Colasante, D.}} \yr{2016} Youtube videos. Available at \url{https://www.youtube.com/watch?v=kG6vXGidQbo} (2016) and \url{https://www.youtube.com/watch?v=ruDpMhlKy6M} (2024). Accessed on April 9, 2025.

\bibitem[Derr {\em et~al.\/}(2022)Derr, Dombrowski, Rycroft \& Klotsa]{Derr2022}
{\sc \au{Derr, N.J}, \au{Dombrowski, T.}, \au{Rycroft, C.H.} \& \au{Klotsa, D.}} \yr{2022}  \at{Reciprocal swimming at intermediate {R}eynolds number}.  \jt{J. Fluid Mech.}  \bvol{952},  \pg{A8}.

\bibitem[Fedder {\em et~al.\/}(2015)Fedder, Hierold, Korvink \& Tabata]{Fedder2015}
{\sc \au{Fedder, G.K.}, \au{Hierold, C.}, \au{Korvink, J.G.} \& \au{Tabata, O.}} \yr{2015} {\em Resonant MEMS: fundamentals, implementation, and application\/}.  \publ{John Wiley \& Sons}.

\bibitem[Fouxon \& Leshansky(2018)]{Fouxon2018}
{\sc \au{Fouxon, I.} \& \au{Leshansky, A.}} \yr{2018}  \at{Fundamental solution of unsteady {S}tokes equations and force on an oscillating sphere near a wall}.  \jt{Phys. Rev. E}  \bvol{98}~(6),  \pg{063108}.

\bibitem[Fouxon {\em et~al.\/}(2020)Fouxon, Rubinstein, Weinstein \& Leshansky]{Fouxon2020}
{\sc \au{Fouxon, I.}, \au{Rubinstein, B.}, \au{Weinstein, O.} \& \au{Leshansky, A.}} \yr{2020}  \at{Fluid-mediated force on a particle due to an oscillating plate and its effect on deposition measurements by a quartz crystal microbalance}.  \jt{Phys. Rev. Lett.}  \bvol{125}~(14),  \pg{144501}.

\bibitem[Gazzola {\em et~al.\/}(2014)Gazzola, Argentina \& Mahadevan]{Gazzola2014}
{\sc \au{Gazzola, M.}, \au{Argentina, M.} \& \au{Mahadevan, L.}} \yr{2014}  \at{Scaling macroscopic aquatic locomotion}.  \jt{Nat. Phys.}  \bvol{10}~(10),  \pg{758--761}.

\bibitem[Hashimoto {\em et~al.\/}(1996)Hashimoto, Koike \& Ueha]{Hashimoto1996}
{\sc \au{Hashimoto, Y.}, \au{Koike, Y.} \& \au{Ueha, S.}} \yr{1996}  \at{Near-field acoustic levitation of planar specimens using flexural vibration}.  \jt{J. Acoust. Soc. Am.}  \bvol{100}~(4),  \pg{2057--2061}.

\bibitem[Hong {\em et~al.\/}(2009)Hong, Liu, Yang \& Zhai]{Hong2009}
{\sc \au{Hong, Q.}, \au{Liu, R.}, \au{Yang, H.} \& \au{Zhai, X.}} \yr{2009} Wall climbing robot enabled by a novel and robust vibration suction technology.  \bt{In {\em 2009 IEEE International Conference on Automation and Logistics\/}},  \pg{pp. 331--336}. IEEE.

\bibitem[Hori(2006)]{Hori2006}
{\sc \au{Hori, Y.}} \yr{2006} {\em Hydrodynamic lubrication\/}.  \publ{Springer}.

\bibitem[Ishizawa(1966)]{Ishizawa1966}
{\sc \au{Ishizawa, S.}} \yr{1966}  \at{The unsteady laminar flow between two parallel discs with arbitrarily varying gap width}.  \jt{Bulletin of JSME}  \bvol{9}~(35),  \pg{533--550}.

\bibitem[Jia {\em et~al.\/}(2023)Jia, Ramanarayanan, Sanchez \& Tolley]{Jia2023}
{\sc \au{Jia, C.}, \au{Ramanarayanan, S.}, \au{Sanchez, A.L.} \& \au{Tolley, M.T.}} \yr{2023}  \at{Controlling the motion of gas-lubricated adhesive disks using multiple vibration sources}.  \jt{Front. Robot. AI.}  \bvol{10},  \pg{1231976}.

\bibitem[Jones \& Wilson(1975)]{Jones1975}
{\sc \au{Jones, A.F.} \& \au{Wilson, S.D.R.}} \yr{1975}  \at{On the failure of lubrication theory in squeezing flows}.  \jt{J. Tribol.} .

\bibitem[Koch {\em et~al.\/}(2021)Koch, Weishaupt, Gl\"aser {\em et~al.\/}]{Koch2021}
{\sc \au{Koch, T.}, \au{Weishaupt, K.}, \au{Gl\"aser, D.} \& \au{others}} \yr{2021}  \at{{DuMux} 3 {\textendash} an open-source simulator for solving flow and transport problems in porous media with a focus on model coupling}.  \jt{Computers \& Mathematics with Applications}  \bvol{81},  \pg{423--443}.

\bibitem[Kuroda \& Hori(1976)]{Kuroda1976}
{\sc \au{Kuroda, S.} \& \au{Hori, Y.}} \yr{1976}  \at{A study of fluid inertia effects in a squeeze film (in japanese)}.  \jt{Journal of Japan Society of Lubrication Engineers}  \bvol{21}~(11),  \pg{740--747}.

\bibitem[Kuzma(1968)]{Kuzma1968}
{\sc \au{Kuzma, D.C.}} \yr{1968}  \at{Fluid inertia effects in squeeze films}.  \jt{Applied Scientific Research}  \bvol{18},  \pg{15--20}.

\bibitem[Landau \& Lifshitz(1986)]{Landau1986}
{\sc \au{Landau, L.D.} \& \au{Lifshitz, E.M.}} \yr{1986} {\em Course of Theoretical Physics vol. 7: Theory of Elasticity\/}.  \publ{3rd edn. Pergamon}.

\bibitem[Langlois(1962)]{Langlois1962}
{\sc \au{Langlois, W.E.}} \yr{1962}  \at{Isothermal squeeze films}.  \jt{Quarterly of Applied Mathematics}  \bvol{20}~(2),  \pg{131--150}.

\bibitem[Lauga(2007)]{Lauga2007}
{\sc \au{Lauga, E.}} \yr{2007}  \at{Floppy swimming: Viscous locomotion of actuated elastica}.  \jt{Phys. Rev. E}  \bvol{75}~(4),  \pg{041916}.

\bibitem[Lee {\em et~al.\/}(2009)Lee, Tung, Raman, Sumali \& Sullivan]{Lee2009}
{\sc \au{Lee, J.W.}, \au{Tung, R.}, \au{Raman, A.}, \au{Sumali, H.} \& \au{Sullivan, J.P.}} \yr{2009}  \at{Squeeze-film damping of flexible microcantilevers at low ambient pressures: theory and experiment}.  \jt{J. Micromech. Microeng.}  \bvol{19}~(10),  \pg{105029}.

\bibitem[Li {\em et~al.\/}(2015)Li, Li, Tao, Liu \& Kagawa]{Li2015}
{\sc \au{Li, X.}, \au{Li, N.}, \au{Tao, G.}, \au{Liu, H.} \& \au{Kagawa, T.}} \yr{2015}  \at{Experimental comparison of {B}ernoulli gripper and vortex gripper}.  \jt{Int. J. Precis. Eng. Man.}  \bvol{16},  \pg{2081--2090}.

\bibitem[Liu {\em et~al.\/}(2023)Liu, Zhao \& Chen]{Liu2023}
{\sc \au{Liu, Y.}, \au{Zhao, Z.} \& \au{Chen, W.}} \yr{2023}  \at{Theoretical investigation of the levitation force generated by underwater squeeze action}.  \jt{Jpn. J. Appl. Phys.}  \bvol{62}~(3),  \pg{034001}.

\bibitem[Mandre {\em et~al.\/}(2009)Mandre, Mani \& Brenner]{Mandre2009}
{\sc \au{Mandre, S.}, \au{Mani, M.} \& \au{Brenner, M.P.}} \yr{2009}  \at{Precursors to splashing of liquid droplets on a solid surface}.  \jt{Phys. Rev. Lett.}  \bvol{102}~(13),  \pg{134502}.

\bibitem[Melikhov {\em et~al.\/}(2016)Melikhov, Chivilikhin, Amosov \& Jeanson]{Melikhov2016}
{\sc \au{Melikhov, I.}, \au{Chivilikhin, S.}, \au{Amosov, A.} \& \au{Jeanson, R.}} \yr{2016}  \at{Viscoacoustic model for near-field ultrasonic levitation}.  \jt{Phys. Rev. E}  \bvol{94}~(5),  \pg{053103}.

\bibitem[Minikes \& Bucher(2003)]{Minikes2003}
{\sc \au{Minikes, A.} \& \au{Bucher, I.}} \yr{2003}  \at{Coupled dynamics of a squeeze-film levitated mass and a vibrating piezoelectric disc: numerical analysis and experimental study}.  \jt{J. Sound Vib.}  \bvol{263}~(2),  \pg{241--268}.

\bibitem[Moore(1965)]{Moore1965}
{\sc \au{Moore, D.F.}} \yr{1965}  \at{A review of squeeze films}.  \jt{Wear}  \bvol{8}~(4),  \pg{245--263}.

\bibitem[Naghdi(1973)]{Naghdi1973}
{\sc \au{Naghdi, P.~M.}} \yr{1973}  \at{The theory of shells and plates}.  \bt{In {\em Linear theories of elasticity and thermoelasticity: linear and nonlinear theories of rods, plates, and shells\/}},  \pg{pp. 425--640}.  \publ{Springer}.

\bibitem[Ollila(1980)]{Ollila1998}
{\sc \au{Ollila, R.G.}} \yr{1980}  \at{Historical review of {WIG} vehicles}.  \jt{J. Hydronaut.}  \bvol{14}~(3),  \pg{65--76}.

\bibitem[Pandey \& Pratap(2007)]{Pandey2007}
{\sc \au{Pandey, A.K.} \& \au{Pratap, R.}} \yr{2007}  \at{Effect of flexural modes on squeeze film damping in {MEMS} cantilever resonators}.  \jt{J. Micromech. Microeng.}  \bvol{17}~(12),  \pg{2475}.

\bibitem[Peng {\em et~al.\/}(2023)Peng, Cuttle, MacMinn \& Pihler-Puzovi{\'c}]{Peng2023}
{\sc \au{Peng, G.G.}, \au{Cuttle, C.}, \au{MacMinn, C.W.} \& \au{Pihler-Puzovi{\'c}, D.}} \yr{2023}  \at{Axisymmetric gas--liquid displacement flow under a confined elastic slab}.  \jt{Phys. Rev. Fluids}  \bvol{8}~(9),  \pg{094005}.

\bibitem[Poulain {\em et~al.\/}(2025)Poulain, Koch, Mahadevan \& Carlson]{Poulain2024}
{\sc \au{Poulain, S.}, \au{Koch, T.}, \au{Mahadevan, L.} \& \au{Carlson, A.}} \yr{2025} Hovering of an actively driven fluid-lubricated foil,  \arxiv{arXiv: 2501.17080}.

\bibitem[Pratap \& Roychowdhury(2014)]{Pratap2014}
{\sc \au{Pratap, R.} \& \au{Roychowdhury, A.}} \yr{2014} {\em Vibratory {MEMS} and Squeeze Film Effects\/},  \pg{pp. 319--338}.  \publ{New Delhi: Springer India}.

\bibitem[Purcell(1977)]{Purcell1977}
{\sc \au{Purcell, E.M.}} \yr{1977}  \at{Life at low {R}eynolds number}.  \jt{Am. J. Phys.}  \bvol{45}~(1),  \pg{3--11}.

\bibitem[Rallabandi(2024)]{Rallabandi2024}
{\sc \au{Rallabandi, B.}} \yr{2024}  \at{Fluid-elastic interactions near contact at low {R}eynolds number}.  \jt{Annu. Rev. Fluid Mech.}  \bvol{56},  \pg{491--519}.

\bibitem[Ramanarayanan(2024)]{ramanarayanan2024emergence}
{\sc \au{Ramanarayanan, S.}} \yr{2024}  \at{On the emergence of attractive load-bearing forces in vibration-induced squeeze-film gas lubrication}. {PhD} thesis, University of California, San Diego.

\bibitem[Ramanarayanan {\em et~al.\/}(2022)Ramanarayanan, Coenen \& S{\'a}nchez]{Ramanarayanan2022}
{\sc \au{Ramanarayanan, S.}, \au{Coenen, W.} \& \au{S{\'a}nchez, A.L.}} \yr{2022}  \at{Viscoacoustic squeeze-film force on a rigid disk undergoing small axial oscillations}.  \jt{J. Fluid Mech.}  \bvol{933}.

\bibitem[Ramanarayanan \& S{\'a}nchez(2022)]{Ramanarayanan2022b}
{\sc \au{Ramanarayanan, S.} \& \au{S{\'a}nchez, A.L.}} \yr{2022}  \at{On the enhanced attractive load capacity of resonant flexural squeeze-film levitators}.  \jt{AIP Advances}  \bvol{12}~(10).

\bibitem[Ramanarayanan \& S{\'a}nchez(2024)]{Ramanarayanan2024}
{\sc \au{Ramanarayanan, S.} \& \au{S{\'a}nchez, A.L.}} \yr{2024}  \at{The role of fluid--structure coupling in the generation of an attractive squeeze-film force}.  \jt{J. Fluid Mech.}  \bvol{1001},  \pg{A52}.

\bibitem[Rayner(1991)]{Rayner1991}
{\sc \au{Rayner, J.M.V.}} \yr{1991}  \at{On the aerodynamics of animal flight in ground effect}.  \jt{Proc. Roy. Soc. B}  \bvol{334}~(1269),  \pg{119--128}.

\bibitem[Rojas {\em et~al.\/}(2010)Rojas, Argentina, Cerda \& Tirapegui]{Rojas2010}
{\sc \au{Rojas, N.O.}, \au{Argentina, M.}, \au{Cerda, E.} \& \au{Tirapegui, E.}} \yr{2010}  \at{Inertial lubrication theory}.  \jt{Phys. Rev. Lett.}  \bvol{104}~(18),  \pg{187801}.

\bibitem[Shi {\em et~al.\/}(2019)Shi, Feng, Hu, Zhu \& Cui]{Shi2019}
{\sc \au{Shi, M.}, \au{Feng, K.}, \au{Hu, J.}, \au{Zhu, J.} \& \au{Cui, H.}} \yr{2019}  \at{Near-field acoustic levitation and applications to bearings: a critical review}.  \jt{Int. J. Extreme Manuf.}  \bvol{1}~(3),  \pg{032002}.

\bibitem[Shintake {\em et~al.\/}(2018)Shintake, Cacucciolo, Floreano \& Shea]{Shintake2018}
{\sc \au{Shintake, J.}, \au{Cacucciolo, V.}, \au{Floreano, D.} \& \au{Shea, H.}} \yr{2018}  \at{Soft robotic grippers}.  \jt{Adv. Mater.}  \bvol{30}~(29),  \pg{1707035}.

\bibitem[Skotheim \& Mahadevan(2005)]{Skotheim2005}
{\sc \au{Skotheim, J.M.} \& \au{Mahadevan, L.}} \yr{2005}  \at{Soft lubrication: The elastohydrodynamics of nonconforming and conforming contacts}.  \jt{Phys. Fluids}  \bvol{17}~(9).

\bibitem[Takasaki {\em et~al.\/}(2010)Takasaki, Terada, Kato, Ishino \& Mizuno]{Takasaki2010}
{\sc \au{Takasaki, M.}, \au{Terada, D.}, \au{Kato, Y.}, \au{Ishino, Y.} \& \au{Mizuno, T.}} \yr{2010}  \at{Non-contact ultrasonic support of minute objects}.  \jt{Phys. Procedia}  \bvol{3}~(1),  \pg{1059--1065}.

\bibitem[Taylor(1967)]{TaylorMovie}
{\sc \au{Taylor, G.I.}} \yr{1967} Film notes for low-{R}eynolds-number flows. National Committee for Fluid Mechanics Films. Available at \url{https://web.mit.edu/hml/ncfmf.html}. Accessed on June 11, 2025.

\bibitem[Taylor \& Saffman(1957)]{Taylor1957}
{\sc \au{Taylor, S.G.} \& \au{Saffman, P.G.}} \yr{1957}  \at{Effects of compressibility at low {R}eynolds number}.  \jt{Journal of the Aeronautical Sciences}  \bvol{24}~(8),  \pg{553--562}.

\bibitem[Tichy \& Bourgin(1985)]{Tichy1985}
{\sc \au{Tichy, J.A.} \& \au{Bourgin, P.}} \yr{1985}  \at{The effect of inertia in lubrication flow including entrance and initial conditions}.  \jt{J. Appl. Mech.} .

\bibitem[Tichy \& Winer(1970)]{Tichy1970}
{\sc \au{Tichy, J.A.} \& \au{Winer, W.O.}} \yr{1970}  \at{Inertial considerations in parallel circular squeeze film bearings}.  \jt{J. Lubric. Tech.} .

\bibitem[Timoshenko \& Woinowsky-Krieger(1959)]{Timoshenko1959}
{\sc \au{Timoshenko, S.} \& \au{Woinowsky-Krieger, S.}} \yr{1959} {\em Theory of plates and shells\/}, 2nd edn.  \publ{McGraw-hill New York}.

\bibitem[Tiwari \& Persson(2019)]{Tiwari2019}
{\sc \au{Tiwari, A.} \& \au{Persson, B.N.J.}} \yr{2019}  \at{Physics of suction cups}.  \jt{Soft matter}  \bvol{15}~(46),  \pg{9482--9499}.

\bibitem[Tuck \& Bentwich(1983)]{Tuck1983}
{\sc \au{Tuck, E.O.} \& \au{Bentwich, M.}} \yr{1983}  \at{Sliding sheets: lubrication with comparable viscous and inertia forces}.  \jt{J. Fluid Mech.}  \bvol{135},  \pg{51--69}.

\bibitem[Ueha {\em et~al.\/}(2000)Ueha, Hashimoto \& Koike]{Ueha2000}
{\sc \au{Ueha, S.}, \au{Hashimoto, Y.} \& \au{Koike, Y.}} \yr{2000}  \at{Non-contact transportation using near-field acoustic levitation}.  \jt{Ultrasonics}  \bvol{38}~(1-8),  \pg{26--32}.

\bibitem[Vandaele {\em et~al.\/}(2005)Vandaele, Lambert \& Delchambre]{Vandaele2005}
{\sc \au{Vandaele, V.}, \au{Lambert, P.} \& \au{Delchambre, A.}} \yr{2005}  \at{Non-contact handling in microassembly: Acoustical levitation}.  \jt{Precis. Eng.}  \bvol{29}~(4),  \pg{491--505}.

\bibitem[Veijola(2004)]{Veijola2004}
{\sc \au{Veijola, T.}} \yr{2004}  \at{Compact models for squeezed-film dampers with inertial and rarefied gas effects}.  \jt{J. Micromech. Microeng.}  \bvol{14}~(7),  \pg{1109}.

\bibitem[Waltham {\em et~al.\/}(2003)Waltham, Bendall \& Kotlicki]{Waltham2003}
{\sc \au{Waltham, C.}, \au{Bendall, S.} \& \au{Kotlicki, A.}} \yr{2003}  \at{Bernoulli levitation}.  \jt{Am. J. Phys.}  \bvol{71}~(2),  \pg{176--179}.

\bibitem[Wei {\em et~al.\/}(2021)Wei, Liu, Zheng, Sun \& Wei]{Wei2021}
{\sc \au{Wei, Z.}, \au{Liu, J.}, \au{Zheng, X.}, \au{Sun, Y.} \& \au{Wei, R.}} \yr{2021}  \at{Influence of squeeze film damping on quality factor in tapping mode atomic force microscope}.  \jt{J. Sound Vib.}  \bvol{491},  \pg{115720}.

\bibitem[Weston-Dawkes {\em et~al.\/}(2021)Weston-Dawkes, Adibnazari, Hu, Everman, Gravish \& Tolley]{Weston2021}
{\sc \au{Weston-Dawkes, W.P.}, \au{Adibnazari, I.}, \au{Hu, Y.-W.}, \au{Everman, M.}, \au{Gravish, N.} \& \au{Tolley, M.T.}} \yr{2021}  \at{Gas-lubricated vibration-based adhesion for robotics}.  \jt{Adv. Intell. Syst.}  \bvol{3}~(7),  \pg{2100001}.

\bibitem[Whitesides(2018)]{Whitesides2019}
{\sc \au{Whitesides, G.M.}} \yr{2018}  \at{Soft robotics}.  \jt{Angew. Chem. Int. Ed.}  \bvol{57}~(16),  \pg{4258--4273}.

\bibitem[Wiertlewski {\em et~al.\/}(2016)Wiertlewski, Fenton~Friesen \& Colgate]{Wiertlewski2016}
{\sc \au{Wiertlewski, M.}, \au{Fenton~Friesen, R.} \& \au{Colgate, J.E.}} \yr{2016}  \at{Partial squeeze film levitation modulates fingertip friction}.  \jt{Proc. Natl. Acad. Sci. U.S.A.}  \bvol{113}~(33),  \pg{9210--9215}.

\bibitem[Wiggins \& Goldstein(1998)]{Wiggins1998b}
{\sc \au{Wiggins, C.H.} \& \au{Goldstein, R.E.}} \yr{1998}  \at{Flexive and propulsive dynamics of elastica at low {R}eynolds number}.  \jt{Phys. Rev. Lett.}  \bvol{80}~(17),  \pg{3879}.

\bibitem[Wiggins {\em et~al.\/}(1998)Wiggins, Riveline, Ott \& Goldstein]{Wiggins1998a}
{\sc \au{Wiggins, C.H.}, \au{Riveline, D.}, \au{Ott, A.} \& \au{Goldstein, R.E.}} \yr{1998}  \at{Trapping and wiggling: elastohydrodynamics of driven microfilaments}.  \jt{Biophys. J.}  \bvol{74}~(2),  \pg{1043--1060}.

\bibitem[Womersley(1955)]{Womersley1955}
{\sc \au{Womersley, J.~R.}} \yr{1955}  \at{Method for the calculation of velocity, rate of flow and viscous drag in arteries when the pressure gradient is known}.  \jt{J. Physiol.}  \bvol{127}~(3),  \pg{553--563}.

\bibitem[Yu {\em et~al.\/}(2006)Yu, Lauga \& Hosoi]{Yu2006}
{\sc \au{Yu, T.S.}, \au{Lauga, E.} \& \au{Hosoi, A.E.}} \yr{2006}  \at{Experimental investigations of elastic tail propulsion at low reynolds number}.  \jt{Phys. Fluids}  \bvol{18}~(9).

\bibitem[Zhang {\em et~al.\/}(2023)Zhang, Bertin, Essink, Zhang, Fares, Shen, Bickel, Salez \& Maali]{Zhang2023}
{\sc \au{Zhang, Z.}, \au{Bertin, V.}, \au{Essink, M.H.}, \au{Zhang, H.}, \au{Fares, N.}, \au{Shen, Z.}, \au{Bickel, T.}, \au{Salez, T.} \& \au{Maali, A.}} \yr{2023}  \at{Unsteady drag force on an immersed sphere oscillating near a wall}.  \jt{J. Fluid Mech.}  \bvol{977},  \pg{A21}.

\bibitem[Zhu {\em et~al.\/}(2006)Zhu, Liu, Wang \& Wang]{Zhu2006}
{\sc \au{Zhu, T.}, \au{Liu, R.}, \au{Wang, X.~D.} \& \au{Wang, K.}} \yr{2006} Principle and application of vibrating suction method.  \bt{In {\em 2006 IEEE International Conference on Robotics and Biomimetics\/}},  \pg{pp. 491--495}. IEEE.

\bibitem[Çengel \& Cimbala(2013)]{Cengel2013}
{\sc \au{Çengel, Y.} \& \au{Cimbala, J.}} \yr{2013} {\em Fluid Mechanics\/}.  \publ{McGraw Hill}.

\end{thebibliography}

\end{document}